\documentclass[preprint,12pt]{elsarticle}

\usepackage{amssymb}
\usepackage{float}
\usepackage{amsmath}

\usepackage{url}
\usepackage{graphicx} % Required for inserting images
\usepackage{epsfig}
\usepackage[dvipsnames]{xcolor}
\usepackage{wrapfig}
\usepackage{longtable}
\usepackage[utf8]{inputenc}
\usepackage{textgreek}
\usepackage{caption}
\usepackage{natbib}

\journal{Spectra Chimica Acta B}

\begin{document}

\begin{frontmatter}


%% Article title
\title{Signatures of rare-earth elements in mineralogical form using laser-ablation dual-comb spectroscopy}

%% Author name
\author[QMI,UBC]{Christina Hofer\corref{cor1}}
\author[QMI,UBC]{Errol Bowman}
\author[Opt]{Andrew Jarymowycz}
\author[Opt]{John J. McCauley}
\author[Opt,UofA]{Dylan Tooley}
\author[Opt]{Hope Dannar}
\author[QMI]{Avery Wong}
\author[QMI]{Ian Pang}
\author[QMI,UBC]{Arthur K. Mills}
\author[Opt]{Mark Phillips}
\author[Opt]{R. Jason Jones}
\author[QMI,UBC]{David J. Jones}

\cortext[cor1]{e-mail: christina.hofer@ubc.ca}

\affiliation[QMI]{organization={Quantum Matter Institute, University of British Columbia},%Department and Organization
            addressline={2355 East Mall}, 
            city={Vancouver},
            postcode={V6T 1Z4}, 
            state={BC},
            country={Canada}}

\affiliation[UBC]{organization={Department of Physics and Astronomy, University of British Columbia},%Department and Organization
            addressline={6224 Agricultural Road}, 
            city={Vancouver},
            postcode={V6T 1Z1}, 
            state={BC},
            country={Canada}}

\affiliation[Opt]{organization={Wyant College of Optical Sciences, University of Arizona},%Department and Organization
            addressline={1630 E. University Blvd.}, 
            city={Tucson},
            postcode={85721}, 
            state={AZ},
            country={USA}}

\affiliation[UofA]{organization={Department of Physics, University of Arizona},%Department and Organization
            addressline={1118 E. Fourth Street}, 
            city={Tucson},
            postcode={85721}, 
            state={AZ},
            country={USA}}

%document
\begin{abstract}
Spectroscopy of laser-produced plasmas offers an avenue for real-time, standoff and non-preparatory sensing capability of rare-earth elements (REEs) within a mineralogical context with applications spanning exploration geology to ore body mapping to ore sorting. Demonstrations of laser-induced breakdown spectroscopy (LIBS) in rock samples have employed both atomic and molecular detection for REE sensors. In this work we evaluate a complementary spectroscopy technique of absorption spectroscopy realized with dual-frequency combs. As this approach provides multi-THz (nm) spectral coverage with simultaneous sub-GHz (pm) resolution, it is ideal for multi-species evaluations present within mineralogical samples because it can improve accuracy and line identification confidence in congested multi-species spectra, which is challenging for emission LIBS. To that end, we analyze REE signatures in calibrated reference materials (CRMs) and a synthesized, REE-containing alloy for atomic, ionic and molecular (oxide) absorptions across three different spectral windows. We identify lines from rare-earth and matrix elements, compare absorption line strengths between the different spectral windows and species and investigate their temporal evolution. For La I, Sm I and Ce I, preliminary limits of detection from 54-583 ppm are estimated for CRMs, using univariate analysis of selected transitions. Comparing the CRM signatures to those of REEs synthesized in a copper alloy, we observe that all REE lines appear earlier and disappear faster in the CRM samples. We attribute these dynamics to matrix effects: Among other elements, the increased oxygen content in the CRM could favor molecular formation. For actual rock samples, observations will once again differ due to grain sizes and bonding mechanisms. Compared to LIBS, we are able to resolve the individual REE and matrix lines with minimal spectral overlap. These proof-of-principle results form a foundation for further development of this laser-based method as a mining sensor.    
\end{abstract}

%% Keywords

\end{frontmatter}

\section{Introduction}
Rare-earth elements (REEs) are a subset of a larger collection of elements (Li, Co, Cu and others) and other compounds (such as Fluorspar, Graphite, Potash, etc.) that are termed ``critical minerals'' \cite{InteriorDepartmentReleases, canadaCriticalMineralsOpportunity2022} as they play crucial roles in sustainable, low-carbon energy sources and the digital economy. In order to mine and process these minerals in a more efficient and environmentally responsible manner, sensitive and selective detectors with real-time capability are crucial to streamline mineral exploration, excavation, and processing. State-of-the-art techniques for real-time mineralogical characterization include magnetic resonance (MR) \cite{coghillBulkSortingTrial2024}, hyperspectral imaging \cite{dalmDiscriminatingOreWaste2017,jobRealtimeShovelMounted2017}, Fourier-transform infrared spectroscopy (FTIR) \cite{dehaineGeometallurgicalCharacterisationPortable2022,destaUseRGBImaging2017}, and Raman spectroscopy \cite{uusitaloOnlineAnalysisMinerals2020, potgieter-vermaakRamanSpectroscopyAnalysis2011}. For elemental composition analysis, commonly employed methods include prompt gamma neutron activation analysis (PGNAA) \cite{abdelnourPromptGammaNeutron2025}, pulsed fast thermal neutron activation (PFTNA) \cite{depPulsedFastThermal1997}, and x-ray fluorescence (XRF) \cite{cetinDeploymentXRFSensors2023, niemelaRealtimeMaterialFlow2015}.

Due to their chemical similarity, the quantification of REEs has proven complicated, particularly for mixtures of several of them. For most analytical techniques, this results in interferences and coincidences \cite{zawiszaDeterminationRareEarth2011}. Inductively-coupled plasma mass spectrometry (ICP-MS) and (high-resolution) continuum wave source electrothermal atomic absorption spectrometry (HR-CS-ETAAS) \cite{welzHighResolutionContinuumSource2005} have proven to be the most sensitive techniques for trace-element (ppb) level REE detection \cite{zawiszaDeterminationRareEarth2011}. HR-CS-ETAAS measures broadband sample absorption after thermal excitation. Both methods can detect multiple REEs in complex samples, with disadvantages in sample preparation requirements and remaining occasional spectral interferences. Avoiding sample preparation, laser-ablation ICP-MS (LA-ICP-MS) uses laser ablation for micro-sampling a specimen \cite{jarvisLaserAblationInductively1993}. Tradeoffs with all these techniques span several considerations such as volumetric vs. surface sampling, safety issues (radiation and radioactive aspects), sample preparation requirements, time scales of obtaining measurement results as well as specificity and detectability/sensitivity of different elements in each technique.\newline

Laser-induced breakdown spectroscopy (LIBS) \cite{liRealTimeHighprecision2023, porterMinExCRCLIBS, harmonLaserInducedBreakdownSpectroscopy2019} employs laser-produced plasmas (LPPs) to detect atoms, ions, molecules, or in other words \textit{analytes} \cite{harilalSpectroscopicCharacterizationLaserinduced2005,harilalOpticalSpectroscopyLaserproduced2018d, harilalOpticalDiagnosticsLaserproduced2022} in the sample under study. LIBS does not require sample preparation, has relatively low technological overhead and demonstrated deployment in other industrial settings \cite{legnaioliIndustrialApplicationsLaserinduced2020}. Therefore, interest in LIBS as a mining sensor has been steadily gaining traction. Recently, the chemical information obtained from LIBS has also been used for mineralogy \cite{harmonLaserInducedBreakdownSpectroscopy2019, diasCalibratingHandheldLIBS2023}. However, the complex and congested spectra from mineral samples, particularly with respect to REEs, pose a distinct challenge for LIBS to untangle \cite{bhattDeterminationRareEarth2018} (see Fig. \ref{fig:LIBSvsDCS} for an example). The relatively large atomic number and special electronic configuration of the REEs means that they have more or denser emission/absorption lines in the UV to visible ranges compared to other/lighter species. The resolution difficulties of LIBS can be traced back to two well-known limitations, that are both fundamental and technical in nature. On one hand, LIBS is an emission-based technique, where lines are observed early after ablation when the plasma temperature is high and line-broadening mechanisms (Stark, Doppler) and self-absorption effects are substantial.  On the other hand, commonly-used, compact grating spectrometers can be limiting the resolution. For the latter, the highest performing LIBS devices typically achieve a resolution around $10\,\mathrm{pm}$ with Czerny-Turner spectrographs with path lengths on the order of meters \cite{zhangEchelleGratingSpectroscopic2022}. Echelle spectrometers have achieved several pm resolution with propagation lengths below a meter \cite{cremersMonitoringUraniumHydrogen2012}. It should be noted that despite both limitations, application of sophisticated data analysis techniques have obtained encouraging results \cite{diazLaserinducedBreakdownSpectroscopy2026}. In addition, interferences have been tackled by probing the plasma at late delays and with a combination of complementary techniques \cite{gaftImagingRareearthElements2019a}.\newline

Significant improvements to the resolution of spectral signatures can be made when analyzing the absorption spectrum of LPP's \cite{harilalOpticalDiagnosticsLaserproduced2022, mertenLaserablationAbsorptionSpectroscopy2022}, because this allows probing when the plasma is no longer emitting, i.e. it has cooled down and broadening effects are less severe. Moreover, this approach can {\it directly} measure the analytes' column densities in the LPP. Tunable, continuous wave (CW) laser absorption spectroscopy (LAS) exploits a narrowband CW source whose linewidth enables optical resolution well below typical LPP absorption linewidths. When combined with the ability to probe a cooler plasma at later times, this improved resolution (compared to LIBS) can parse a more congested optical spectrum and collect continuous time delays \cite{harilalTimeresolvedAbsorptionSpectroscopic2021, walaCharacterizationElectronDensity2025}. However, the spectral range of high-resolutions scans for tunable CW lasers is typically limited to $<60\,\mathrm{GHz}$ ($100\,\mathrm{pm}$). While multiple scan regions can be acquired over a larger overall tuning range \cite{phillipsGroundStateRotational2025}, this approach increases measurement acquisition times. As a result, CW-laser-based LAS has found most success targeting relatively narrow absorption windows and therefore one or two elemental species \cite{harilalOpticalDiagnosticsLaserproduced2022} at a time.\newline 

Dual-comb spectroscopy (DCS) \cite{coddingtonDualcombSpectroscopy2016} is particularly suited to overcome resolution limitations while also covering multi-THz spectral bandwidths as coveted in mineralogical analysis for multi-species identification. DCS builds on recent technological advances in frequency comb metrology and features sub-GHz resolution, multi-THz optical bandwidth and $\mathrm{\mu s}$-to-ms measurement acquisition times. DCS has recently been applied to measure absorption of LPPs \cite{bergevinDualcombSpectroscopyLaserinduced2018b,zhangTimeresolvedDualcombMeasurement2019}, and allows one to simultaneously track temporal dynamics for a variety of species of interest with absorption lines from the ultraviolet to the infrared. Flexibility regarding the spectral coverage can be achieved through harmonic and super-continuum generation \cite{muravievDualfrequencycombUVSpectroscopy2024,camenzindUltralowNoiseSpectral2025, mashburnVisibleDualcombSpectroscopy2026}. Recent demonstrations have also shown the capability of dual-comb sensors to be field-deployed and miniaturized \cite{hermanPreciseMultispeciesAgricultural2021, zhongBroadbandPhotoncountingDualcomb2025, duttOnchipDualcombSource2018}.\newline

As an absorption-based method, DCS can measure LPP spectra under lower temperature conditions than LIBS and thus can access narrow-linewidth, low and ground state initial energy ($E_i$) lines. So far, one or two individual REEs have been measured with DCS in metallic alloys with low or controlled oxygen content \cite{weeksMeasurementNeutralGadolinium2021a, weeksMultispeciesTemperatureNumber2022c, rhoadesDualcombAbsorptionSpectroscopy2022}. In realistic mining samples, REEs are present in mineral structures that vary between lithologies. This surrounding material is what is called the mineral \textit{matrix}. For example, Ce, La and Nd are often found in their oxide states CeO$_2$, La$_2$O$_3$, and Nd$_2$O$_3$ within carbonatite-derived ores, where the surrounding lateritic matrix also has a high concentration of iron and aluminum oxides. In addition to the large number of REE lines in the UV and visible mentioned above, (strong) lines from the mineral matrix can interfere with the REE signatures.\newline

Different matrices are expected to have distinct ablation and LPP properties and therefore affect absorption spectra and temporal evolution. Laser-ablation DCS (LA-DCS) in particular is capable of measuring these dynamics for three main reasons: (i) high resolution to distinguish many narrow lines; (ii) broad-enough optical bandwidth to cover entire molecular branches, and (iii) it can measure a wide range of delays ($\mu$s to ms) after ablation to track the temporal evolution.\newline

Building on previous results for alloys containing 1 or 2 REEs \cite{weeksMeasurementNeutralGadolinium2021a, weeksMultispeciesTemperatureNumber2022c, rhoadesDualcombAbsorptionSpectroscopy2022}, in this work we present the first LA-DCS signatures of REEs in calibrated reference materials (CRM), which are  fully assayed (or characterized) ore samples in a powder form. We identify matrix and REE absorption lines and investigate the temporal evolution of absorption lines of REEs in their atomic, ionic and oxide states in a LPP after ablation. We also compare the observed lines and temporal behavior to that of three REEs in a copper alloy. We showcase the ability of LA-DCS to detect various lines of multiple REEs in both samples and investigate the effects of the differing matrices. These measurements represent a first step towards the application of LA-DCS for the detection of REEs in real-world mineral samples \cite{hoferDualCombSpectroscopyRareEarthElementDetection2025, bowmanDualCombSpectroscopySystem2025}.\newline

The following methods section gives an overview of the experimental setup and measurement principle, as well as the sample preparation. In the results section, we look at absorption lines of atomic, ionic, and molecular nature in the blue, NIR, and green spectral range. In addition, we show preliminary limits of detection (LOD) for several REE species and compare LIBS and DCS spectra. We then discuss the experimental results and interpret the observed dynamics. The paper ends with conclusions and an outlook to future experiments and applications of LA-DCS in critical mineral detection.\newline

\section{Methods}
\subsection{Experimental setup}\label{sec:setup}

\begin{figure}[t]
    \centering \includegraphics{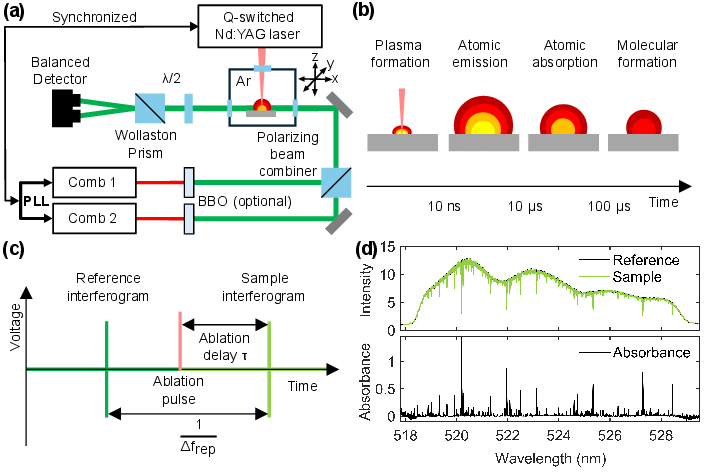}
    \caption{Experimental setup and measurement principle. (a) Schematic of the experimental setup. Two frequency combs are synchronized via phase-locked loops (PLL), the comb timing is also synchronized with the data acquisition and laser ablation (details in text). The combs are either used directly for dual-comb spectroscopy of the laser-produced plasma (LPP) or frequency doubled in BBO crystals before probing the sample. A balanced detector measures the generated interferograms after transmission of both beams through the LPP. The sample can be translated in all three dimensions to allow for scanning during ablation and for adjusting the probe height. (b) Temporal evolution of the LPP after ablation and dominant processes for spectroscopy. (c) Measurement timing sequence. Interferograms occur at a rate of $\Delta f_\mathrm{rep}$. A reference interferogram is recorded before ablation and the ablation event is set to happen at a delay $\tau$ before the recorded sample interferogram. (d) Exemplary resulting spectral information. The reference and sample spectra are the result of a phase-corrected Fourier transform of the interferograms. Taking the logarithmic ratio of the two spectra results in the plotted absorbance.}
    \label{fig:setup}
\end{figure}

The data presented in this paper was collected with two different LA-DCS setups, all similar to the one described in Ref. \cite{weeksMeasurementNeutralGadolinium2021a}. Figure \ref{fig:setup}(a) shows a schematic of the general experimental layout. Specifics for each laser source will be given at the end of this section.

In short, the DCS measurement principle follows that of Fourier-transform spectroscopy, with frequency combs used as the light source. Two of these femtosecond lasers (Comb 1, Comb 2) are phase-coherently locked together using phase-locked loops (PLLs) \cite{tourigny-planteOpenFlexibleDigital2018} and two continous wave (CW) lasers \cite{coddingtonDualcombSpectroscopy2016, weeksMeasurementNeutralGadolinium2021a}. One of the CW lasers is used as a (free-running) optical reference for both combs' spacing, while the second CW laser (spectrally-separated from the first) acts as a transfer oscillator to complete the phase coherence between the two combs. The phase-coherent synchronization ensures stable beating frequencies between the combs, corresponding to a constant repetition rate difference $\Delta f_\mathrm{rep}$. The difference in repetition rate determines the rate at which the laser pulses are swept through each other, resulting in the generation of interferograms. The optical outputs of the two lasers are either directly combined with orthogonal polarizations or first doubled in frequency via second-harmonic generation in a BBO nonlinear crystal. In the Ar-purged sample chamber, a Q-switched Nd:YAG laser ablates a portion of the sample into a laser-produced plasma (LPP). The combined DCS beams probe the generated LPP a few mm above the sample surface. After their interaction with the plasma, the beams are spectrally filtered and pass through a $\lambda /2$ plate and a Wollaston prism (WP) for balanced detection. Spectral filtering serves two purposes: (i) it reduces background from plasma emission; and (ii) it limits the down-converted  radio-frequency spectrum to one free spectral range and thus avoids aliasing \cite{coddingtonDualcombSpectroscopy2016, weeksMeasurementNeutralGadolinium2021a}.\newline

We temporally synchronize the DCS data acquisition and laser ablation by measuring $\Delta f_\mathrm{rep}$ and triggering the ablation at a set time before the interferogram maximum for the sample measurement (Fig. \ref{fig:setup}(c)). In this manner, a reproducible delay $\tau$ between ablation and interferogram peak/acquisition is ensured, allowing several spectra per delay to be averaged and to take a time series of absorption spectra. A reference interferogram is measured before each ablation event. Varying the ablation delay probes different stages of the plasma evolution or varying temperatures (Fig. \ref{fig:setup}(b)) and also varies the population distribution in the lower state energies ($E_i$), ionization fraction and molecular formation. Due to the co-linear alignment of Comb 1 and Comb 2, in which both act as probe beams, the absorption information is symmetrically contained both before and after the interferogram peak. However, the ablation event limits the usable time-domain information before the interferogram. We therefore mirror the information from after the peak in what we term \textit{single-sided analysis} \cite{ben-davidComputationSpectrumSinglebeam2002,mccauleyDualcombSpectroscopyDeep2024a}. With this data-processing method, we can apodize the temporal data with a Blackman-Harris window of identical length for all ablation delays and therefore keep the spectral resolution constant. Averaging the spectra retrieved before and after the ablation pulse results in reference and sample spectra like the ones shown in Fig. \ref{fig:setup}(d) (top). Calculating the natural logarithmic ratio of the two gives the absorbance depicted on the bottom of that panel, with the relative frequency axis determined from $f_\mathrm{rep}$ and $\Delta f_\mathrm{frep}$. The wavelength axis is calibrated by applying a constant offset that matches the measured spectra to known absorption lines. \newline
The ablation event happens with a ns laser in a pressure (and flow)-regulated Ar atmosphere. The ns laser heats the plasma after it is first generated and the Ar environment slows down the plasma evolution compared to vacuum \cite{degiacomoEffectsBackgroundEnvironment2012}, resulting in hot and comparatively long-lived plasmas. Additionally, the Ar flow creates an inert environment to provide reproducible ablation conditions for this initial work. Pressure and flow were adjusted for maximum absorption signatures and will be listed in Section \ref{sec:results}. DCS probe powers below $1\,\mathrm{mW}$ per comb ensure that only the ablation laser excites the plasma and that the photodiodes are not saturated.\newline

We use three different sources to probe with various wavelength DCS beams:

\begin{itemize}
    \item Near-infrared (NIR): The fundamental wavelengths of two custom-built Ti:Sapphire lasers operating near $800\,\mathrm{nm}$.
    \item Green: The second harmonic wavelengths of two commercial Yb:fiber lasers, around $520\,\mathrm{nm}$. 
    \item Blue: The second harmonics of the Ti:Sapphire lasers (mentioned above for the NIR probe), near $400\,\mathrm{nm}$.
\end{itemize}

In all three spectral regions, we currently have the capability to cover a $\approx 10\,\mathrm{nm}$ (or $4-20\,\mathrm{THz}$, depending on the central wavelength) wide wavelength window. These spectral ranges cover several, strong absorption lines for each REE species. As noted above, the 10 nm width is not a fundamental limit, as supercontinuum-based comb sources can significantly expand the range. Current research focuses on broadening the coverage in the visible \cite{mashburnVisibleDualcombSpectroscopy2026} and UV \cite{mccauleyDualcombSpectroscopyDeep2024a}. This would allow accessing spectral regions with even stronger lines. Because of the high density of REE lines in the UV and visible, we pick representative smaller regions for the data shown in Section \ref{sec:results} in the blue and green spectral ranges. This choice highlights the high resolution and remaining high number of distinguishable spectral features.\newline 

For each absorbance spectrum, neutral and ionized atomic absorption lines are identified using absorption line positions from the Kurucz data base \cite{kuruczAtomicLineData1995} matched to the NIST spectral line data base \cite{AtomicSpectraDatabase2009} within $2\,\mathrm{pm}$. Building on REE transitions previously observed with DCS, we select those data base lines with initial state energies $E_\mathrm{i}\leq15000\,\mathrm{cm^{-1}}$. We expect lines with a reported oscillator strength of log(gf) $\leq -1.5$ to be very weak, but acknowledge uncertainties in these values. To identify observed Ce I-lines not present in the Kurucz and NIST data base in the green spectral range, we also calculate the dipole-allowed transitions from the energy levels in the NIST data base \cite{AtomicSpectraDatabase2009, weeksMeasurementNeutralGadolinium2021a}. This calculation selects those optical transitions which fulfill the selection rules for the quantum numbers of the involved energy levels, and combined with the high spectral resolution of DCS provides high-confidence line assignment to a set of energy levels for a given element, although it does not provide oscillator strength information \cite{weeksMeasurementNeutralGadolinium2021a}. We use spectral modeling data available from the ExoMol database and simulate molecular spectra in the software PGOPHER to determine rotational-vibrational electronic band locations for LaO \cite{westernPGOPHERProgramSimulating2017,bernathStypeStarsLine2023}. For matching of lines from the Kurucz and NIST databases and for initial guesses for the line identification in the experimental data, we used Anthropic. (2026). Claude (Opus 4.8) [large language model]. We subsequently checked all the experimental line matches manually. 

\subsection{Samples and sample preparation}\label{sec:samples}

We investigate the absorption features of REEs in two different matrices. The first is a synthesized alloy consisting of a copper base with specific amounts of added REEs and the second is a set of CRMs prepared from source materials that contain the oxide-states of REEs, mineralized as carbonatite in a weathered, lateritic matrix (Fe and Al rich). Both samples are assayed and certified and are described in detail below.\newline

\begin{enumerate}

    \item Metal alloy: Calibrated sputtering target from ACI Alloys, Inc. \cite{SputteringTargetsEvaporation} with Cu ($90.0\pm 0.1\%$), La ($4.0\pm 0.1\%$), Ce ($4.0\pm 0.1\%$), Nd ($2.0\pm 0.1\%$), where the REEs are present in a metallic alloy with low oxygen content.

    \item CRM: A set of REE CRMs containing 100 certified values (and 50 indicative values) for element concentrations (\cite{REEs}, certificates and characterization methods available on the OREAS website). The homogenized material contains waste, low and medium REE ores. The raw materials were dried, roasted, crushed and milled before they were assayed and blended to achieve the desired grades. Table \ref{tab:dilution_samples} summarizes the CRMs and their REE concentrations. In all cases, the CRM powders are pressed into pellets using KBr powder as a binder \cite{gondalRoleVariousBinding2007a} with a ratio of 1:3. After this mixture is ground with mortar and pestle to a grain size of $<5\,\mathrm{\mu m}$, approximately $1.5\,\mathrm{g}$ of material are pressed in a $13\,\mathrm{mm}$-diameter die with a $1.5\,\mathrm{t}$ press for $1\,\mathrm{min}$. Sample 1 (OREAS 465 with the binder) is used for studies in Section \ref{sec:REEsignatures} while the dilution series of section \ref{sec:LOD} employs samples 1-8. Note that samples 7 and 8 used an inert ground quartz CRM (OREAS22H \cite{22H}) for further dilution. 
    
\end{enumerate}

\begin{table}[t]
    \centering
    \small
    \caption{Calibrated REE powders and element concentrations for the samples used in the dilution series (\ref{sec:LOD}). For samples 7 and 8, OREAS22H powder is used for dilution of OREAS 460.}
    \begin{tabular}{cccccc}
        \hline
         Sample & CRM & Ce (\%) & La (\%) & Sm (\%) & Nd (\%) \\
         \hline
         1 & 465 & 1.3 & 0.81 & 0.051 & 0.4\\
         2 & 464 & 0.51 & 0.39 & 0.47& 0.3\\
         3 & 463 & 0.22 & 0.17 & 0.017& 0.12\\
         4 & 462 & 0.17 & 0.13 & 0.012& 0.087\\
         5 & 461 & 0.12 & 0.095 & $7.7\times 10^{-5}$& 0.058\\
         6 & 460 & 0.06 & 0.046 & $3.6\times 10^{-5}$& 0.027\\
         7 & 460/2 & 0.030 & 0.023 & $1.8\times 10^{-5}$& 0.014\\
         8 & 460/4 & 0.015 & 0.012 & $8.9\times 10^{-6}$& $6.8\times 10^{-5}$\\
         \hline
    \end{tabular}
    \label{tab:dilution_samples}
\end{table}

\begin{figure}[t]
    \centering
    \includegraphics[width=0.5\linewidth]{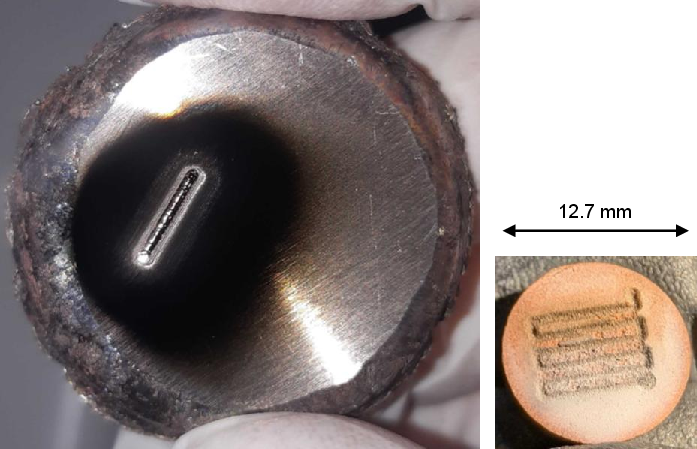}
    \caption{Pictures of ablated alloy (left) and pellet (right) samples. The sample is translated under the ablation laser in order to expose new material for every shot.}
    \label{fig:samples}
\end{figure}

Pictures of two samples are shown in Fig. \ref{fig:samples}.

\section{Results and Discussion}\label{sec:results}

\begin{figure}
    \centering
    \includegraphics[width=1\linewidth]{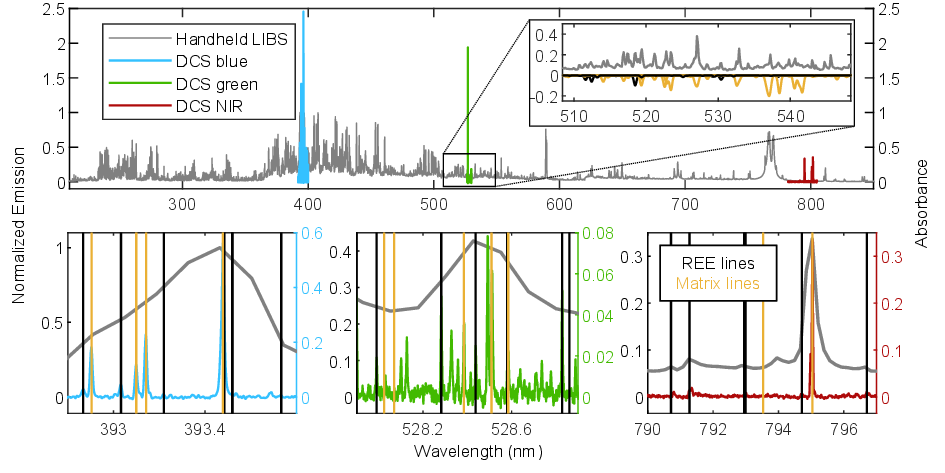}
    \caption{Comparison of LIBS and DCS spectra for a CRM (OREAS465) pressed into a pellet, with top panel displaying the full LIBS spectral coverage and bottom panel zooming in on subsets of the blue, green and NIR regions measured with DCS. Grey line: LIBS data (left y-axis, normalized emission). Colored lines: DCS data for $30\,\mathrm{\mu s}$ delay after the ablation for the three spectral regions (right y-axis: absorbance). Vertical lines: positions of absorption lines from the Kurucz and NIST databases, with REE lines in black and matrix lines in yellow. The inset in the top panel zooms in to a portion of the LIBS spectrum with a similar number of resolved lines as for the DCS spectra on the bottom. The negative axis shows simulations of the LIBS spectra for REEs (black) and matrix elements (yellow).}
    \label{fig:LIBSvsDCS}
\end{figure}

As an example for a typical emission spectrum, a representative LIBS measurement of CRM sample 1 from Table \ref{tab:dilution_samples} is shown in gray in the top panel of Fig. \ref{fig:LIBSvsDCS}. This spectrum was measured with a commercial instrument (SciAps Z-903 Analyzer) with a $50\,\mathrm{mJ}$ ablation pulse energy, $100\,\mathrm{\mu m}$ spot size and a spectral resolution of $0.1-0.2\,\mathrm{nm}$. It was acquired $645\,\mathrm{ns}$ after ablation with a gate width of about $14\,\mathrm{\mu s}$ under atmospheric pressure with Ar flow. The LIBS spectral features are about $0.4-0.6\,\mathrm{nm}$ wide, as shown in the bottom panel. Due to the instrumental resolution, each visible peak in the spectrum may consist of multiple overlapping transitions from one or more REE or matrix elements, making analysis of these complex REE minerals challenging. Nevertheless, recent multivariate analysis of LIBS data on this type of sample has made notable progress \cite{diazLaserinducedBreakdownSpectroscopy2026}. 

Also in the top panel of Fig. \ref{fig:LIBSvsDCS} are the three spectral windows measured in this DCS-based work, chosen for the overlap with transitions of the ionic, neutral and molecular oxide forms of REEs. The lower panels are zoomed-in regions comparing parts of the LIBS and DCS spectra. Vertical, thin lines indicate absorption line positions from the Kurucz data base for various elements, black for the REEs and yellow for matrix elements. In comparison to LIBS, the lines in the DCS spectrum are less than $0.01\,\mathrm{nm}$ wide. Numerous REE and matrix lines are distinguishable in the DCS spectra where the LIBS spectra only show a shoulder and cannot provide selectivity between elements.

A full analysis of the measured DCS lines in various REE CRMs is discussed below and compared with alloys containing REEs to evaluate matrix effects. 

\subsection{REE signatures in calibrated ore samples and alloys}\label{sec:REEsignatures}
Each absorbance spectrum in this section is the result of averaging spectra from 750 or 1000 ablation laser shots, corresponding to a measurement time of about one minute, limited by the ablation laser repetition rate of 10-20 Hz. We translate the sample under the ablation laser with a speed of $0.2\,\mathrm{mm/s}$ to expose new material for each laser shot. Due to its powder nature, the crater drilling in the pellet sample is deeper than in the alloy. 
Depending on the species, REEs have absorption features in different spectral regions. We cover parts of these spectral regions using the LA-DCS laser systems described in Section \ref{sec:setup} to probe ionic, molecular and neutral lines in the LPP. A sequential discussion of each follows now.\newline

\begin{figure}[h]
    \centering
    \includegraphics[width=1\textwidth]{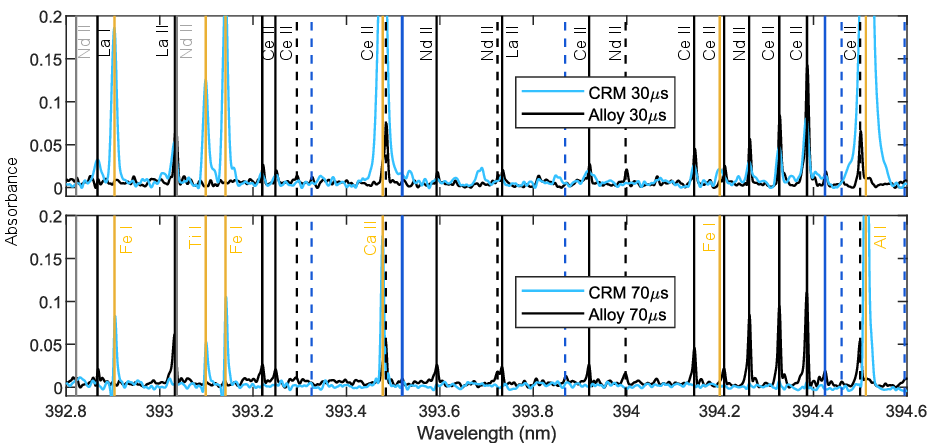}
    \caption{CRM (blue) and alloy (black) absorption spectra in the blue spectral range. Top-panel curves are obtained from interferograms measured $30\,\mathrm{\mu s}$ after the ablation event, bottom-panel curves at $70\,\mathrm{\mu s}$. Vertical lines indicate identified lines from the Kurucz and NIST database: Yellow lines correspond to Fe, Ti, Ca and Al, which are matrix elements present only in the pellet. Black, solid, labeled lines where observed in both CRM and alloy samples, dashed ones only in the alloy. Blue lines are only present in the Kurucz data base. Spectral resolution from apodization window length: $12.5\,\mathrm{GHz}$/$7\,\mathrm{pm}$. A baseline fit has been subtracted from all spectra to cancel out slow fluctuations caused by imbalancing and drifts. The vertical scale was chosen to highlight REE lines. At $30\,\mathbf{\mu s}$, the Ca II and Al I lines have an absorbance of $0.57$ and $1.42$, respectively.}
    \label{fig:alloypelletblue}
\end{figure}

A large number of REE ionic absorption lines are present in the blue spectral range and we focus here on the spectral window between $392.8-394.6\,\mathrm{nm}$ \cite{jarymowyczCombingRareEarths2025}. The repetition rate of the laser oscillators is $f_\mathrm{rep,blue}\approx 179\,\mathrm{MHz}$ with a detuning of $\Delta f_\mathrm{rep, blue}\approx269\,\mathrm{Hz}$. The ablation laser pulse energy was $236\,\mathrm{mJ}$ with a spot size of approximately $1\,\mathrm{mm}$. Absorption features were detected for an Ar pressure of $110\,\mathrm{mbar}$ with a flow rate of $50\,\mathrm{lpm}$ and a probe height $3.5\,\mathrm{mm}$ above the sample surface. The absorption spectra of both samples at early (top) and later (bottom) delays are shown in Fig. \ref{fig:alloypelletblue} for 750 averages each.

The strongest features in the CRM spectrum are from the matrix elements Fe I, Ti I, Ca II and Al I (yellow lines) which - apart from the REEs - are the highest abundance elements in the CRM according to the certification ($29.55\mathrm{wt}.\%$ Fe, $10.51\mathrm{wt}.\%$ TiO$_2$, $0.87\mathrm{wt}.\%$ Ca, $6.21\mathrm{wt}.\%$ Al) \cite{REEs} with absorption lines measurable in our spectral range. At least a factor of 5 weaker, predominantly ionic REE signatures from La, Nd and Ce are visible, some of them in the wings of other lines. All these ionic lines have disappeared by $70\,\mathrm{\mu s}$. In the alloy, the same and more (dashed lines) ionic REE signatures were observed, were initially stronger and lasted longer than in the CRM. In the alloy, some lines are easier to identify without the strong background/matrix features in the CRM (yellow lines). For both samples, the measured REE lines positions match those of the databases within $2.4\,\mathrm{pm}$. Table \ref{tab:blue_lines} summarizes the absorption lines for $wt.\%$-level abundance species present in the Kurucz and NIST data bases and indicates their strength and wavelength accuracy in the two samples.

\begin{table}[ht]
\centering
\caption{Summary of the absorption lines measured in the blue spectral range - see Fig. \ref{fig:alloypelletblue}, matched to the Kurucz and NIST data bases. Peak absorbances are measured at $30\,\mathrm{\mu s}$ for the CRM and $70\,\mathrm{\mu s}$ for the alloy sample. Oscillator strength (log(gf)) and lower energy level ($E_\mathrm{i}$) are taken from the Kurucz data base. $\Delta_\mathrm{CRM}$ and $\Delta_\mathrm{All.}$ show the wavelength mismatch between Kurucz data base and measurement. {\it n/o} indicates the line is not observed and {\it n/a} that this species is not present in the sample. Interfering lines are grouped together with horizontal lines.}
\label{tab:blue_lines}
\small
\begin{tabular}{cccccccc}
\hline
Species & $\lambda_{\mathrm{vac}}$ (nm) & log(gf) & $E_{\mathrm{i}}$ (cm$^{-1}$) & A$_\mathrm{CRM}$ & A$_\mathrm{All.}$ & $\Delta_{\mathrm{CRM}}$ (pm) & $\Delta_{\mathrm{All.}}$ (pm) \\
\hline
Nd II & 392.8210 & -0.550 & 1470.1 & n/o & n/o & - & - \\
La I & 392.8668 & -0.750 & 0.0 & 0.019 & 0.018 & 0.4 & 1.4 \\
Fe I & 392.9032 & -1.590 & 888.1 & 0.186 & n/a & -0.3 & - \\
\hline
La II & 393.0324 & -0.240 & 1394.5 & 0.026 & 0.056 & -0.9 & -0.3 \\
Nd II & 393.0368 & -0.580 & 4512.5 & n/o & n/o & - & -2.0 \\
\hline
Ti I & 393.0987 & -1.060 & 0.0 & 0.126 & n/a & -0.2 & - \\
Fe I & 393.1410 & -1.590 & 704.0 & 0.236 & n/a & 0.0 & - \\
Ce II & 393.2196 & -0.417 & 1410.3 & 0.012 & 0.016 & -2.1 & 0.5 \\
Ce II & 393.2479 & -0.693 & 2382.2 & 0.005 & 0.011 & 2.1 & 0.3 \\
Ce II & 393.2936 & -0.387 & 5942.8 & n/o & 0.090 & - & 0.9 \\
\hline
Ca II & 393.4777 & -0.191 & 0.0 & 0.570 & n/a & 1.0 & - \\
Ce II & 393.4844 & -0.291 & 5675.8 & n/o & 0.055 & - & -0.8 \\
\hline
Nd II & 393.5929 & -0.340 & 2585.5 & 0.010 & 0.040 & 0.1 & 0.8 \\
Nd II & 393.7233 & 0.650 & 12087.2 & n/o & 0.018 & - & 0.5 \\
La II & 393.7332 & -1.340 & 1016.1 & 0.012 & 0.030 & -2.0 & -0.3 \\
Ce II & 393.9195 & 0.831 & 12751.8 & 0.008 & 0.010 & -1.8 & -0.3 \\
Nd II & 393.9978 & 0.870 & 13048.6 & n/o & 0.016 & - & -0.7 \\
Ce II & 394.1445 & -0.280 & 2563.2 & 0.026 & 0.046 & -0.5 & 0.6 \\
Fe I & 394.1993 & -2.600 & 7728.1 & 0.022 & n/a & -0.4 & - \\
Ce II & 394.2086 & -0.577 & 3363.4 & 0.012 & 0.007 & 2.4 & -0.3 \\
Nd II & 394.2623 & -0.280 & 513.3 & 0.012 & 0.068 & 0.0 & 0.7 \\
Ce II & 394.3267 & -0.180 & 0.0 & 0.026 & 0.079 & -0.1 & 0.2 \\
Ce II & 394.3861 & 0.797 & 6913.4 & 0.050 & 0.094 & -1.6 & -0.5 \\
\hline
Ce II & 394.5000 & 0.187 & 6389.9 & n/o & 0.034 & - & -0.4 \\
Al I & 394.5122 & -0.623 & 0.0 & 1.422 & n/a & 1.7 & - \\
\hline
\end{tabular}
\end{table}

\begin{figure} [h]
    \centering
    \includegraphics[width=1\textwidth]{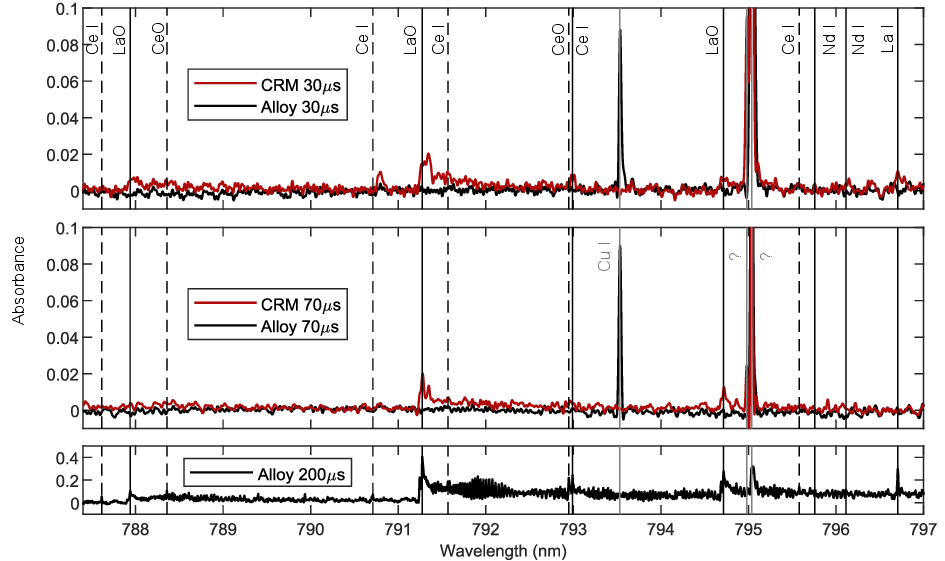}
    \caption{Measured absorbance spectra for CRM (red) and alloy (black) samples in the NIR spectral range corresponding to the molecular absorption lines. The top panel shows pellet data for an ablation delay of $30\,\mathrm{\mu s}$, the middle panel for $70\,\mathrm{\mu s}$. Solid black vertical lines indicate features that were detected in both samples, dashed ones those which were only visible in the alloy. Grey vertical lines are for Cu I and two unknown lines, which are unique to the respective samples. Spectral resolution from apodization window length: $12.3\,\mathrm{GHz}$/$26\,\mathrm{pm}$ for $30\,\mathrm{\mu s}$ and $70\,\mathrm{\mu s}$ delay and $1.4\,\mathrm{GHz}$/$3\,\mathrm{pm}$ for $200\,\mathrm{\mu s}$ delay (recorded with $\Delta f_r\approx800\,\mathrm{Hz}$. A baseline fit has been subtracted from the $30\,\mathrm{\mu s}$ CRM data. The vertical axis limits are chosen to highlight the REE data. At $30\,\mathrm{\mu s}$, the strength of the Ar I line is 0.34 and 0.39 in the CRM and alloy, respectively.}
    \label{fig:pelletdilutionNIR}
\end{figure}

In the CRM - apart from strong matrix lines - 2 Nd, 3 La and 7 Ce lines matching the two data bases were identified.\newline
The fundamental wavelength of the Ti:Sapphire laser system can probe several absorption bands of LaO and CeO in the NIR spectral range \cite{rhoadesDualcombAbsorptionSpectroscopy2022}. We recorded absorption spectra for the CRM and alloy samples with the same conditions as in the blue, see Fig. \ref{fig:pelletdilutionNIR}.

In the CRM, the molecular features are of similar strength at about $70\,\mathrm{\mu s}$ after the ablation event as at $30\,\mathrm{\mu s}$. However, they are relatively weak and only visible for LaO. Looking at further delays beyond  $70\,\mathrm{\mu s}$ we observe a decrease in these signatures. For the alloy, the onset of molecular absorption is further delayed, strong LaO signatures and two distinct CeO features are present at $200\,\mathrm{\mu s}$. Most neutral and ionic transitions in the NIR spectral range are comparatively weak and from highly excited initial energy levels. The strongest neutral Ce and La transitions are marked in Figure \ref{fig:pelletdilutionNIR}, mostly absent in the CRM but detected in the alloy at $200\,\mathrm{\mu s}$. 

\begin{figure}[h]
    \centering
    \includegraphics[width = 1\textwidth]{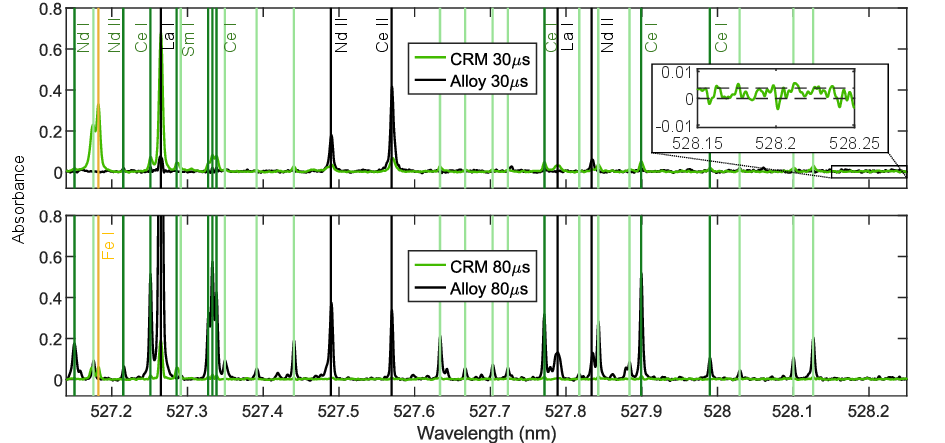}
    \caption{Comparison between LPP absorption spectra after ablation of CRM (green) and alloy (black) in the green spectral range. The top panel shows the spectra for an ablation delay of $30\,\mathrm{\mu s}$, the bottom panel for $80\,\mathrm{\mu s}$. Vertical lines indicate lines from various data-bases: Ce I dipole-allowed transitions in light green, Kurucz database only in dark green, Kurucz and NIST matched in black. A yellow line marks the only matrix line from Fe I. Spectral resolution resulting from apodization window length: about $2.1,\mathrm{GHz}$/$0.002\,\mathrm{nm}$ for all four spectra. The inset in the top right corner shows the noise floor for the $30\,\mathrm{\mu s}$ measurement of the pellet. In this spectral window with no absorption lines the standard deviation of the absorbance is $0.0019$.}
    \label{fig:alloypelletgreen}
\end{figure}

For an example of neutral transitions, we move to the green spectral range and employ two frequency-doubled Yb:fiber lasers. Their average repetition rates are $f_\mathrm{rep, green}\approx100\,\mathrm{MHz}$, with a difference of $\Delta f_\mathrm{rep,green}\approx 790\,\mathrm{Hz}$. We focus on the $527.14-528.25\,\mathrm{nm}$ spectral range, which has a number of lines from various REE species. The ablation laser pulse energy was about $217\,\mathrm{mJ}$ with a spot size of $250\,\mathrm{\mu m}$. The Ar pressure was about $110\,\mathrm{mbar}$ and the probe height $3.5\,\mathrm{mm}$. Figure \ref{fig:alloypelletgreen} shows the CRM (green) and alloy (black) absorbances for $30\,\mathrm{\mu s}$ (top) and $80\,\mathrm{\mu s}$ (bottom) delay.

\begin{figure}[h]
    \centering
    \includegraphics[width=0.48\textwidth]{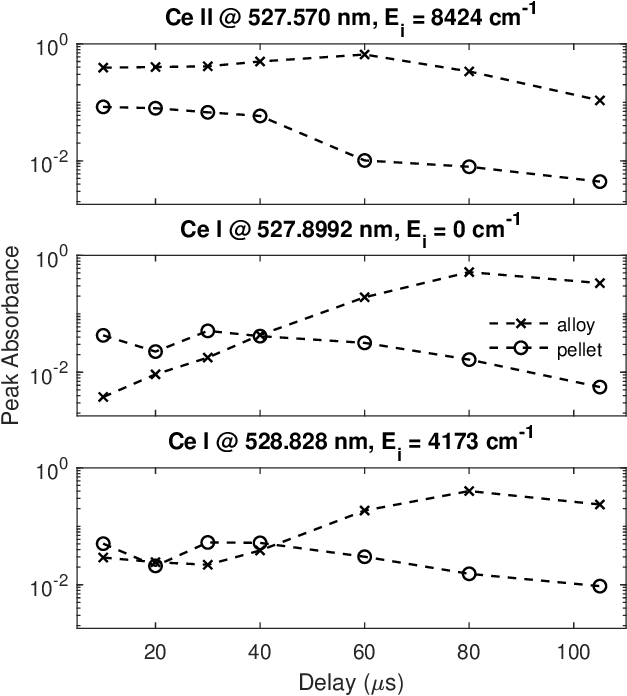}
    \caption{Evolution of peak absorbance as a function of ablation delay for 3 different absorption lines of Ce I and Ce II. The maximum absorbance of each line is determined for various delays, where all data was evaluated with the same, $156\,\mathrm{\mu s}$-wide apodization window.}
    \label{fig:temporalevolution}
\end{figure}

For both samples, we observe several neutral and a few ionic electronic transitions of Nd, La, Sm and Ce. At $30\,\mathrm{\mu s}$, the neutral lines are stronger in the CRM than in the alloy, while the ionic lines dominate the alloy spectrum. In comparison, at $80\,\mathrm{\mu s}$ most REE signatures have disappeared in the CRM and the neutral lines have increased significantly for the alloy. We measured the absorbance for several delays to investigate their time-dependent strength in more detail, see Fig. \ref{fig:temporalevolution}. The Ce I lines are strongest at about $30\,\mathrm{\mu s}$ in the CRM, and at about $80\,\mathrm{\mu s}$ in the alloy, while the Ce II line strength peaks at earlier delays (which informed the delay choices for Fig. \ref{fig:alloypelletgreen}). Similarly to the blue spectral range, data base lines for the most abundant elements are summarized in Table \ref{tab:green_lines}, with absorbance and spectral mismatch values.

\begin{table}[ht]
\centering
\caption{Summary of the absorption lines measured in the blue spectral range - see Fig. \ref{fig:alloypelletgreen}, matched to the Kurucz and NIST data bases. Peak absorbances are measured at $30\,\mathrm{\mu s}$ for the CRM and $80\,\mathrm{\mu s}$ for the alloy sample. Oscillator strength (log(gf)) and lower energy level ($E_\mathrm{i}$) are taken from the Kurucz data base. $\Delta_\mathrm{CRM}$ and $\Delta_\mathrm{All.}$ show the wavelength mismatch between Kurucz data base and measurement. {\it n/a} indicates that this species is not present in the sample and 'sat' that the line was saturated.}
\label{tab:green_lines}
\small
\begin{tabular}{cccccccc}
\hline
Spec. & $\lambda_{\mathrm{vac}}$ (nm) & log(gf) & $E_{\mathrm{i}}$ (cm$^{-1}$) & A$_\mathrm{CRM}$ & A$_\mathrm{All.}$ & $\Delta_{\mathrm{CRM}}$ (pm) & $\Delta_{\mathrm{All.}}$ (pm) \\
\hline
Fe I & 527.1823 & -1.510 & 12968.6 & 0.329 & n/a  & 0.1 & - \\
La I & 527.2649 & -0.790 & 1053.2 & 0.677 & sat. & -0.1 & -0.1 \\
Nd II & 527.4895 & -0.120 & 5487.7 & 0.036 & 0.373 & 0.6 & 0.7 \\
Ce II & 527.5697 & -0.323 & 8423.7 & 0.068 & 0.338 & 1.1 & 0.3 \\
La I & 527.7888 & -0.730 & 8446.0 & 0.132 & 0.031 & -0.6 & 0.1 \\
Nd II & 527.8337 & -0.440 & 6931.8 & 0.012 & 0.129 & 1.3 & 1.5 \\
\hline
\end{tabular}
\end{table}

\begin{table}[ht]
\centering
\caption{Additional absorption lines identified for the green spectrum according to the Kurucz data base, with no matches in the NIST spectral line data base.}
\label{tab:unmatched_green}
\small
\begin{tabular}{cccccccc}
\hline
Spec. & $\lambda_{\mathrm{vac}}$ (nm) & log(gf) & $E_{\mathrm{i}}$ (cm$^{-1}$) & A$_\mathrm{CRM}$ & A$_\mathrm{All.}$ & $\Delta_{\mathrm{CRM}}$ (pm) & $\Delta_{\mathrm{All.}}$ (pm) \\
\hline
Nd I & 527.1508 & 0.300 & 8411.9 & 0.016 & 0.176 & -0.2 & 0.0 \\
Nd II & 527.2153 & -1.630 & 1650.2 & 0.007 & 0.064 & 0.3 & 0.7 \\
Ce I & 527.2511 & -0.529 & 4173.5 & 0.073 & 0.513 & 0.3 & 0.1 \\
Sm I & 527.2857 & -0.305 & 811.9 & 0.045 & n/a & 0.8 & - \\
Ce I & 527.3272 & -0.660 & 3100.2 & 0.044 & 0.325 & 0.3 & 0.7 \\
Ce I & 527.3329 & -0.349 & 4746.6 & 0.082 & 0.574 & 0.1 & 0.0 \\
Ce I & 527.3384 & 0.012 & 8588.0 & 0.077 & 0.0425 & -0.1 & -0.1 \\
Ce I & 527.7714 & -0.084 & 8603.5 & 0.048 & 0.317 & 0.0 & -0.2 \\
Ce I & 527.8992 & -1.355 & 0.0 & 0.051 & 0.517 & -0.0 & 0.2 \\
Ce I & 527.9899 & -0.102 & 10673.8 & 0.020 & 0.105 & 0.4 & 0.1 \\
\hline
\end{tabular}
\end{table}

All the lines from the Kurucz and NIST data base are matched with measured lines within $1.5\,\mathrm{pm}$. In the green spectral range, the NIST data base is more scarcely populated than in the blue, particularly for neutral Ce I lines. We have summarized additional lines matched within $0.8\,\mathrm{pm}$ only to the Kurucz data base in Table \ref{tab:unmatched_green}. Remarkably, in this spectral range all REE lines were observed in both the alloy and CRM with pm-level wavelength precision. Combining both tables, a total of 2 La, 4 Nd and 8 Ce lines were matched. The measured absorbances are at least an order of magnitude above the measurement noise floor (see Section \ref{sec:LOD} below).\newline

For comparable concentrations, the neutral absorption lines in the CRM sample are generally shorter-lived and weaker than in the copper alloy (see section \ref{sec:samples} for the sample composition). Furthermore, ionic lines appear stronger and longer-lasting in the alloy compared to the CRM. Long-lasting excited ionic transitions have been observed previously in absorption spectra of LPPs using both DCS \cite{mccauleyDualcombSpectroscopyDeep2024a} and tunable CW laser spectroscopy \cite{walaCharacterizationElectronDensity2025} and allow for a tracking of the evolution of the ionization fraction. These absorption-based investigations also measured late-time ionization fractions significantly above Saha equilibrium predit. \newline
We attest additional weakening and faster dynamics of the REE features in the CRM to the changed matrix, which has both chemical and mechanical/physical effects. Perhaps most prominently, the increased oxygen content may play a role in molecular oxide formation kinetics that differ for neutral and ionic reactants. Mechanically, the different matrix can also affect the ablation process itself and therefore plasma composition and temperature. Faster dynamics could be attested to a more rapidly expanding and therefore cooling plasma. CRMs and alloys have different thermal conductivity, which changes the crater formation and thus also affects plasma plume conditions. Thus, the initial plasma plume physical conditions (temperature, spatial profile, ionization, etc.) are likely different between the CRM and alloy samples, in turn affecting the dynamics of observed species (ions, neutrals, molecules) and corresponding observed spectral lines.\newline

Observing signatures from REEs in both CRM and alloy samples in three different spectral ranges allows us to identify regions that are promising for detecting REEs or for collecting information about the sample matrix. Prominent REE signatures are observed in the green, where several neutral or ionic absorption lines with large oscillator strengths are present for each species. The blue spectral range shows strong signatures from the Fe, Ti, Al and Ca matrix. In addition, this is where more ionic features of REE can be found. Two weak LaO bands were observed in the NIR, while CeO was not detected in the CRM sample. In the pellet, Ce is present as CeO$_2$ and La as La$_2$O$_3$. Obtaining a more complete understanding of their high-temperature oxide formation would require measuring the higher-order oxides - for example with more broadband spectral coverage in the green spectral range for CeO$_2$ \cite{weltnerSpectroscopyRareEarth1971}.\newline

\subsection{CRM dilution series - limit of detection (LOD)}\label{sec:LOD}
An exemplary noise floor for the LA-DCS measurements is shown in the inset in Fig. \ref{fig:alloypelletgreen}. It shows a spectral region with no lines, where the absorbance has a standard deviation of $0.0019$. This results in peak line strength to noise floor ratios in the CRM sample as high as $358$. The data acquisition noise floor is made up of detector noise and optical shot noise from balanced detection \cite{newburySensitivityCoherentDualcomb2010}. After the calculation of the absorbance, the baseline noise increases where reference and signal spectrum approach this noise floor. Scaling of that baseline standard deviation with the square root of the number of measurements has been shown previously for unstructured noise \cite{weeksDualCombSpectroscopyLaserProduced2023}.\newline

\begin{figure} [h]
    \centering
    \includegraphics[width = 0.8\textwidth]{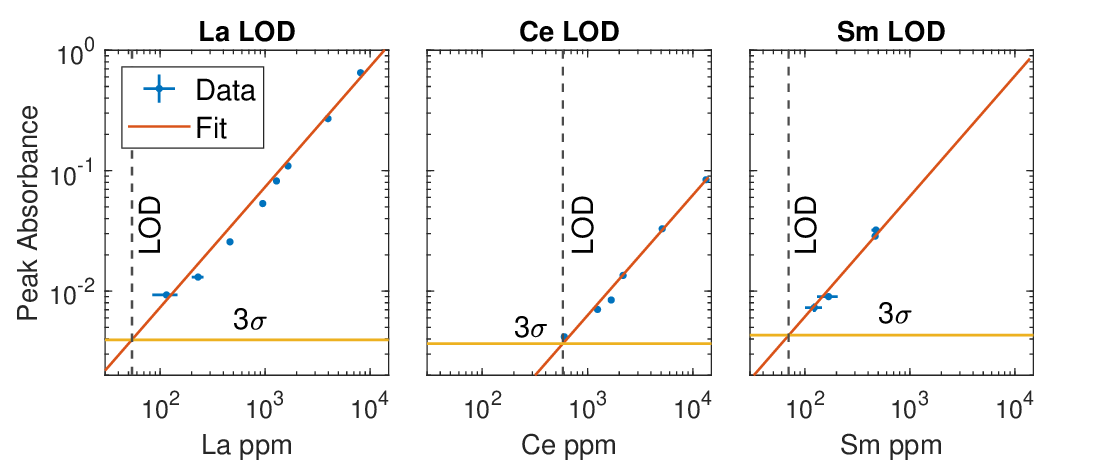}
    \caption{Dilution Series. Peak absorbance as a function of nominal REE concentration (see Table \ref{tab:dilution_samples} for the values). For Ce I and La I data points are shown for all 8 samples, for Sm I, only the results for samples 1-4 are shown, as the absorption line is otherwise too weak to be identified.}
    \label{fig:dilution_LOD}
\end{figure}

The high SNR from the CRM measurement facilitates a dilution series with decreasing REE concentrations to determine a LOD. Here, we measured absorbance spectra for 8 different samples (see Table \ref{tab:dilution_samples}) with decreasing REE concentrations in the green spectral range at $30\,\mathrm{\mu s}$ delay. We picked one strong and isolated line for La, Ce and Sm respectively and tracked its peak absorbance, see Fig. \ref{fig:dilution_LOD}.

The absorbance values follow the same linear trend over 1-2 orders of magnitude for the three different species (linear fit with $0$ intercept). The LOD is defined as in \cite{phillipsDetectionLimitsLaser2025}, namely as the intercept of this linear fit to the data points with the $3\sigma$-line, where $\sigma$ is the standard deviation of the absorbance for a CRM of pure quartz (OREAS 22H \cite{22H}) sample (average of 1000 spectra) in the 0.03-nm-wide range around the spectral position of the absorption line. The $3\sigma$-line for each REE varies due to minor changes in the averaged 22H-absorbance depending on the spectral position. Table \ref{tab:LOD} summarizes the resulting detection limits, which are between $54$ and $583\,\mathrm{ppm}$ for the three different species. The LOD values depend on the strength of the absorption lines, which vary with species (atomic or ionic), ground state energy and oscillator strength. In addition, changes in matrix - for example oxygen content - can differently affect the number of available absorbers per species and therefore also the line strength. As a result, LOD values represent a first demonstration with several possible avenues for improvement. \newline

For this first application of DCS to the detection of REEs in mineral matrices, the observed LOD values are about one order of magnitude higher than the natural abundances of REEs in the earth's crustor than what has been detectable using lab-scale LIBS setups \cite{labutinNovelApproachSensitivity2016,diazLaserinducedBreakdownSpectroscopy2026}. The values are several orders of magnitude above trace-element detection levels with LA-ICP-MS or HR-CS-ETAAS \cite{zawiszaDeterminationRareEarth2011}, but recent efforts have shown dramatic advances in LODs for tunable laser absorption spectroscopy in LPPs \cite{phillipsDetectionLimitsLaser2025}. However, Figures \ref{fig:alloypelletblue} to \ref{fig:alloypelletgreen} showcase the number of absorption lines observable and resolvable with DCS, enabling high-confidence identification of several REEs within a complex matrix. As mentioned above, these first LOD estimates used a single, not necessarily ideal line per element. Particularly for Ce, stronger, neutral lines are present in other spectral regions. Advancing the data analysis, particularly employing the full number of detectable lines using multivariate techniques which are established in LIBS \cite{bhattDeterminationRareEarth2018,martinQuantificationRareEarth2015}, promises significant improvements in LOD. Furthermore, according to the noise-floor-scaling described in Refs. \cite{weeksDualCombSpectroscopyLaserProduced2023, phillipsDetectionLimitsLaser2025}, increasing the number of measurements or the interaction length with the plasma could further increase the sensitivity.

\begin{table}[t]
\caption{Summary of LODs for La, Ce and Sm, using just one absorption line for each element.}
    \centering
    \small
    \begin{tabular}{c c c c c}
    \hline
      Species & $\lambda_\mathrm{vac}$ (nm) & LOD (ppm by mass) & $E_i (\mathrm{cm^{-1}})$ & log(gf) \\
      \hline
      La I & 527.265 & $54\pm 0.05$ & 1053.2 & -0.79 \\
      Ce II & 527.570 & $583\pm 3$ & 8423.7 & -0.323 \\
      Sm I & 527.285 & $70\pm 0.8$ & 811.9 & -0.305 \\
      \hline
    \end{tabular}
    \label{tab:LOD}
\end{table} 

\section{Conclusion and Outlook}
In this work, we have presented the first LA-DCS measurements of REE CRMs and compared them to measurements of REEs in copper alloys. These proof-of-principle measurements identified absorption signatures of REEs in their neutral, ionic, and molecular form in various spectral regions with excellent agreement with database agreement regarding the transition wavelengths. For each species, several absorption lines were matched to the Kurucz and NIST databases, setting the basis for multivariate element identification. Several REEs were measured simultaneously, showcasing the capabilities of DCS thanks to its high resolution. Almost all database lines were distinguishable, with few spectral interferences remaining for very strong matrix lines. Our measurements also showcased differences occurring due to the change in matrix between elemental and mineral samples. The presence of several distinct and strong features in the green for multiple species allows for concentration retrieval and plasma temperature determination using a Boltzmann distribution \cite{weeksMultispeciesTemperatureNumber2022c}. With this number density analysis, LA-DCS is able to determine relative concentrations in the LPP and could therefore potentially perform mineralogical identification. In addition, information about the matrix can be obtained, e.g. from the Fe I and Ti I lines in the blue spectral range, which are separable from the REE because of the excellent resolution. A more detailed study of the plasma dynamics could be complemented with the oxide information from the NIR. All this information combined would result in a more quantitative study of the plasma creation and evolution affected by the different matrices, which will be the focus of future studies.\newline

The CRM measurements are a first step towards detecting REEs in natural, heterogeneous ore material. Future work will move towards solid rock samples and investigate the detection limits for the various species of interest in those real-world samples. In addition, the plasma dynamics can be investigated in more detail, particularly by studying the plasma temperature, ionization fractions, and chemical composition as a function of time after ablation. Such studies could give insights behind the weakening of the absorption lines compared to alloy samples and the early disappearance.\newline

Through broadening in nonlinear fibers, LA-DCS has the capability to cover similar spectral ranges to LIBS, however with orders of magnitude improved resolution. This would allow to cover all spectral ranges discussed here with a single laser system. In addition, it enables DCS to also detect Li and Co, which are both crucial for battery manufacturing and other renewable energy sources. These elements have absorption lines in the visible and UV. As it is a very light element, Li cannot be detected with XRF measurements, but is an excellent candidate for optical absorption spectroscopy, which has shown low LODs \cite{phillipsDetectionLimitsLaser2025} and the ability to measure $^6$Li/$^7$Li isotope ratios with high precision and accuracy \cite{phillips2026CalibrationFree}. On the other hand, Co is not bound to oxygen in its mineral form, which might result in more straightforward plasma dynamics and detection. Co is present in waste rocks from Ni and Cu mines \cite{zhangStirredtankBioleachingCopper2022,saimComprehensiveReviewCobalt2024}, where the detection of small concentrations on a background of various elements could exploit the benefits of DCS. The heterogeneous nature of REE-bearing rocks (both primary and in tailings) requires high-density characterization to ensure these elements are representatively detected. In a mining environment, the ability to rapidly and accurately characterize any REEs present provides an opportunity to recover these critical minerals and prevent them from being discarded or lost.\newline
Conveyor-belt implementation and competitive detection limits for LA-DCS still require several steps of technological advancement. Nevertheless, with its current capabilities it promises to serve as a calibration tool for LIBS and as a drill-core analyzer. 

\section{Acknowledgments}
The authors thank Cassedy Harraden for helpful discussions, input to the final manuscript and for assisting with the handheld LIBS measurements.\newline
This research was undertaken thanks in part to funding from the Canada First Research Excellence Fund, Quantum Materials and Future Technologies. We also acknowledge the support of the Natural Sciences and Engineering Research Council of Canada (RGPIN-2020-07085, ALLRP 586674-2023), New Frontiers in Research Fund NFRFE-2020-00626, Canada Foundation for Innovation (CFI) Project 4366, and the British Columbia Knowledge Development Fund (BCKDF). RJJ acknowledges support from AFOSR (FA9550-20-1-0273)

\section{Data availability}
Data will be made available upon reasonable request.

\bibliographystyle{elsarticle-num-names}
\bibliography{REE_Paper_review_noURL}

@article{tourigny-planteOpenFlexibleDigital2018,
	title = {An open and flexible digital phase-locked loop for optical metrology},
	volume = {89},
	issn = {0034-6748},
	doi = {10.1063/1.5039344},
	number = {9},
	urldate = {2026-05-15},
	journal = {Rev. Sci. Instrum.},
	author = {Tourigny-Plante, Alex and Michaud-Belleau, Vincent and Bourbeau Hébert, Nicolas and Bergeron, Hugo and Genest, Jérôme and Deschênes, Jean-Daniel},
	month = sep,
	year = {2018},
	pages = {093103},
}

@article{saimComprehensiveReviewCobalt2024,
	title = {A {Comprehensive} {Review} on {Cobalt} {Bioleaching} from {Primary} and {Tailings} {Sources}},
	volume = {45},
	issn = {0882-7508},
	doi = {10.1080/08827508.2023.2181346},
	number = {5},
	urldate = {2026-04-20},
	journal = {Mineral Processing and Extractive Metallurgy Review},
	publisher = {Taylor \& Francis},
	author = {Saim, Alex Kwasi and Darteh, Francis Kwaku},
	month = jul,
	year = {2024},
	pages = {426--452},
}

@article{zhangStirredtankBioleachingCopper2022,
	title = {Stirred-tank bioleaching of copper and cobalt from mine tailings in {Chile}},
	volume = {180},
	issn = {0892-6875},
	doi = {10.1016/j.mineng.2022.107514},
	urldate = {2026-04-20},
	journal = {Minerals Engineering},
	author = {Zhang, Ruiyong and Schippers, Axel},
	month = apr,
	year = {2022},
	pages = {107514},
}

@article{ben-davidComputationSpectrumSinglebeam2002,
	title = {Computation of a spectrum from a single-beam {Fourier}-transform infrared interferogram},
	volume = {41},
	copyright = {© 2002 Optical Society of America},
	issn = {2155-3165},
	doi = {10.1364/AO.41.001181},
	language = {EN},
	number = {6},
	urldate = {2026-04-20},
	journal = {Appl. Opt., AO},
	publisher = {Optica Publishing Group},
	author = {Ben-David, Avishai and Ifarraguerri, Agustin},
	month = feb,
	year = {2002},
	pages = {1181--1189},
}

@article{diasCalibratingHandheldLIBS2023,
	title = {Calibrating a {Handheld} {LIBS} for {Li} {Exploration} in the {Barroso}–{Alvão} {Aplite}-{Pegmatite} {Field}, {Northern} {Portugal}: {Textural} {Precautions} and {Procedures} {When} {Analyzing} {Spodumene} and {Petalite}},
	volume = {13},
	copyright = {http://creativecommons.org/licenses/by/3.0/},
	issn = {2075-163X},
	shorttitle = {Calibrating a {Handheld} {LIBS} for {Li} {Exploration} in the {Barroso}–{Alvão} {Aplite}-{Pegmatite} {Field}, {Northern} {Portugal}},
	doi = {10.3390/min13040470},
	language = {en},
	number = {4},
	urldate = {2026-04-20},
	journal = {Minerals},
	publisher = {Multidisciplinary Digital Publishing Institute},
	author = {Dias, Filipa and Ribeiro, Ricardo and Gonçalves, Filipe and Lima, Alexandre and Roda-Robles, Encarnación and Martins, Tânia},
	month = apr,
	year = {2023},
	pages = {470},
}

@misc{mashburnVisibleDualcombSpectroscopy2026,
	title = {Visible dual-comb spectroscopy across more than 100 {THz} with lithium niobate nanophotonic waveguides},
	doi = {10.48550/arXiv.2602.07239},
	urldate = {2026-03-27},
	publisher = {arXiv},
	author = {Mashburn, Carter and Chang, Kristina F. and Wahl, Michael J. and Walsh, Mathieu and Herman, Daniel I. and Heyrich, Matthew and Wu, Tsung-Han and Hoghooghi, Nazanin and Sekine, Ryoto and Ledezma, Luis and Jerris, Emily and Marandi, Alireza and Genest, Jerome and Diddams, Scott A.},
	month = feb,
	year = {2026},
}

@article{phillipsGroundStateRotational2025,
	title = {Ground state rotational and kinetic temperatures of {C2} molecules in a laser-produced plasma},
	volume = {138},
	issn = {0021-8979},
	doi = {10.1063/5.0283658},
	number = {5},
	urldate = {2026-02-17},
	journal = {J. Appl. Phys.},
	author = {Phillips, Mark C. and Harilal, Sivanandan S.},
	month = aug,
	year = {2025},
	pages = {053302},
}

@article{harilalTimeresolvedAbsorptionSpectroscopic2021,
	title = {Time-resolved absorption spectroscopic characterization of ultrafast laser-produced plasmas under varying background pressures},
	volume = {103},
	doi = {10.1103/PhysRevE.103.013213},
	number = {1},
	urldate = {2026-02-17},
	journal = {Phys. Rev. E},
	publisher = {American Physical Society},
	author = {Harilal, S. S. and Kautz, E. J. and Phillips, M. C.},
	month = jan,
	year = {2021},
	pages = {013213},
}

@article{walaCharacterizationElectronDensity2025,
	title = {Characterization of electron density and ionization of a uranium laser produced plasma using laser absorption spectroscopy},
	volume = {227},
	issn = {0584-8547},
	doi = {10.1016/j.sab.2025.107142},
	urldate = {2026-02-17},
	journal = {Spectrochimica Acta Part B: Atomic Spectroscopy},
	author = {Wala, Ryland G. and Polek, Mathew P. and Harilal, Sivanandan S. and Jones, R. Jason and Phillips, Mark C.},
	month = may,
	year = {2025},
	pages = {107142},
}

@article{mertenLaserablationAbsorptionSpectroscopy2022,
	title = {Laser-ablation absorption spectroscopy: {Reviewing} an uncommon hyphenation},
	volume = {189},
	issn = {0584-8547},
	shorttitle = {Laser-ablation absorption spectroscopy},
	doi = {10.1016/j.sab.2022.106358},
	urldate = {2026-02-17},
	journal = {Spectrochimica Acta Part B: Atomic Spectroscopy},
	author = {Merten, Jonathan},
	month = mar,
	year = {2022},
	pages = {106358},
}

@article{cremersMonitoringUraniumHydrogen2012,
	title = {Monitoring {Uranium}, {Hydrogen}, and {Lithium} and {Their} {Isotopes} {Using} a {Compact} {Laser}-{Induced} {Breakdown} {Spectroscopy} ({LIBS}) {Probe} and {High}-{Resolution} {Spectrometer}},
	volume = {66},
	issn = {0003-7028},
	doi = {10.1366/11-06314},
	language = {EN},
	number = {3},
	urldate = {2026-02-17},
	journal = {Appl Spectrosc},
	publisher = {SAGE Publications Ltd STM},
	author = {Cremers, David A. and Beddingfield, Alan and Smithwick, Robert and Chinni, Rosemarie C. and Jones, C. Randy and Beardsley, Burt and Karch, Larry},
	month = mar,
	year = {2012},
	pages = {250--261},
}

@article{camenzindUltralowNoiseSpectral2025,
	title = {Ultra-low noise spectral broadening of two combs in a single {ANDi} fiber},
	volume = {10},
	issn = {2378-0967},
	doi = {10.1063/5.0251190},
	number = {3},
	urldate = {2026-02-04},
	journal = {APL Photonics},
	author = {Camenzind, Sandro L. and Sierro, Benoît and Willenberg, Benjamin and Nussbaum-Lapping, Alexander and Rampur, Anupamaa and Keller, Ursula and Heidt, Alexander M. and Phillips, Christopher R.},
	month = mar,
	year = {2025},
	pages = {036119},
}

@article{weltnerSpectroscopyRareEarth1971,
	title = {Spectroscopy of rare earth oxide molecules in inert matrices at 4.deg.{K}},
	volume = {75},
	issn = {0022-3654, 1541-5740},
	doi = {10.1021/j100674a013},
	language = {en},
	number = {4},
	urldate = {2026-02-03},
	journal = {J. Phys. Chem.},
	author = {Weltner, William and DeKock, Roger L.},
	month = feb,
	year = {1971},
	pages = {514--525},
}

@article{labutinNovelApproachSensitivity2016,
	title = {A novel approach to sensitivity evaluation of laser-induced breakdown spectroscopy for rare earth elements determination},
	volume = {31},
	issn = {1364-5544},
	doi = {10.1039/C6JA00200E},
	language = {en},
	number = {11},
	urldate = {2026-01-29},
	journal = {J. Anal. At. Spectrom.},
	publisher = {The Royal Society of Chemistry},
	author = {Labutin, Timur A. and Zaytsev, Sergey M. and Popov, Andrey M. and Zorov, Nikita B.},
	month = oct,
	year = {2016},
	pages = {2223--2226},
}

@misc{SputteringTargetsEvaporation,
	title = {Sputtering {Targets}, {Evaporation} \& {Thin} {Film} {Materials} {\textbar} {High} {Purity}, {Catalyst} \& {Precious} {Metal} {Alloys} {\textbar} {ACI} {ALLOYS}, {INC}},
	language = {en-US},
    url = {https://www.acialloys.com/},
	year = {Accessed: 2026-01-24},
}

@article{jarvisLaserAblationInductively1993,
	title = {Laser ablation inductively coupled plasma mass spectrometry ({LA}-{ICP}-{MS}): a rapid technique for the direct, quantitative determination of major, trace and rare-earth elements in geological samples},
	volume = {106},
	issn = {0009-2541},
	shorttitle = {Laser ablation inductively coupled plasma mass spectrometry ({LA}-{ICP}-{MS})},
	doi = {10.1016/0009-2541(93)90030-M},
	number = {3},
	urldate = {2026-01-23},
	journal = {Chemical Geology},
	author = {Jarvis, Kym E. and Williams, John G.},
	month = jun,
	year = {1993},
	pages = {251--262},
}

@article{legnaioliIndustrialApplicationsLaserinduced2020,
	title = {Industrial applications of laser-induced breakdown spectroscopy: a review},
	volume = {12},
	issn = {1759-9679},
	shorttitle = {Industrial applications of laser-induced breakdown spectroscopy},
	doi = {10.1039/C9AY02728A},
	language = {en},
	number = {8},
	urldate = {2026-01-23},
	journal = {Anal. Methods},
	publisher = {The Royal Society of Chemistry},
	author = {Legnaioli, S. and Campanella, B. and Poggialini, F. and Pagnotta, S. and Harith, M. A. and Abdel-Salam, Z. A. and Palleschi, V.},
	month = feb,
	year = {2020},
	pages = {1014--1029},
}

@article{harmonLaserInducedBreakdownSpectroscopy2019,
	title = {Laser-{Induced} {Breakdown} {Spectroscopy}—{An} {Emerging} {Analytical} {Tool} for {Mineral} {Exploration}},
	volume = {9},
	copyright = {http://creativecommons.org/licenses/by/3.0/},
	issn = {2075-163X},
	doi = {10.3390/min9120718},
	language = {en},
	number = {12},
	urldate = {2026-01-23},
	journal = {Minerals},
	publisher = {Multidisciplinary Digital Publishing Institute},
	author = {Harmon, Russell S. and Lawley, Christopher J. M. and Watts, Jordan and Harraden, Cassady L. and Somers, Andrew M. and Hark, Richard R.},
	month = dec,
	year = {2019},
	pages = {718},
}

@misc{22H,
	title = {{22H}},
	language = {en-AU},
	urldate = {2026-01-23},
	journal = {OREAS},
    url = {https://www.oreas.com/crm/oreas-22h/},
    year = {Accessed: 2026-01-23},
}

@article{zhangEchelleGratingSpectroscopic2022,
	title = {Echelle {Grating} {Spectroscopic} {Technology} for {High}-{Resolution} and {Broadband} {Spectral} {Measurement}},
	volume = {12},
	copyright = {http://creativecommons.org/licenses/by/3.0/},
	issn = {2076-3417},
	doi = {10.3390/app122111042},
	language = {en},
	number = {21},
	urldate = {2026-01-23},
	journal = {Applied Sciences},
	publisher = {Multidisciplinary Digital Publishing Institute},
	author = {Zhang, Yinxin and Li, Wanzhuo and Duan, Wenhao and Huang, Zhanhua and Yang, Huaidong},
	month = jan,
	year = {2022},
	pages = {11042},
}

@misc{REEs,
	title = {{REEs}},
	language = {en-AU},
    url = {https://www.oreas.com/search/?newGroup=REE},
    year = {Accessed: 2026-01-22},
	journal = {OREAS},
}

@article{porterMinExCRCLIBS,
	title = {{MinEx} {CRC} {LIBS} {Downhole} {Geochemistry} and {EM} {Swept}-frequency {Tools} {Feature} in {Coring} {Magazine} {MinEx} {CRC} {Technologies} {Feature} in the {April}/{May} {Edition} of {Australasian} {Drilling} {Magazine}},
	language = {en},
	author = {Porter, Anna},
    year = {2025},
}

@article{liRealTimeHighprecision2023,
	title = {Real time and high-precision online determination of main components in iron ore using spectral refinement algorithm based {LIBS}},
	volume = {31},
	copyright = {© 2023 Optica Publishing Group},
	issn = {1094-4087},
	doi = {10.1364/OE.505574},
	language = {EN},
	number = {23},
	urldate = {2026-01-22},
	journal = {Opt. Express, OE},
	publisher = {Optica Publishing Group},
	author = {Li, An and Zhang, Xinyu and Liu, Xiaodong and He, Yage and Shan, Yuheng and Sun, Haohan and Yi, Wen and Liu, Ruibin},
	month = nov,
	year = {2023},
	pages = {38728--38743},
}

@article{cetinDeploymentXRFSensors2023,
	title = {Deployment of {XRF} {Sensors} {Underground}: {An} {Opportunity} for {Grade} {Monitoring} or {Bulk} {Ore} {Sorting} in {Cave} {Mines}},
	volume = {13},
	copyright = {http://creativecommons.org/licenses/by/3.0/},
	issn = {2075-163X},
	shorttitle = {Deployment of {XRF} {Sensors} {Underground}},
	doi = {10.3390/min13050672},
	language = {en},
	number = {5},
	urldate = {2026-01-22},
	journal = {Minerals},
	publisher = {Multidisciplinary Digital Publishing Institute},
	author = {Cetin, Mahir Can and Klein, Bern and Li, Genzhuang and Futcher, William and Haest, Maarten and Welsh, Andrew},
	month = may,
	year = {2023},
	pages = {672},
}

@article{depPulsedFastThermal1997,
	title = {Pulsed fast and thermal neutron analysis for coal and cement industries},
	volume = {392},
	issn = {0094-243X},
	doi = {10.1063/1.52722},
	number = {1},
	urldate = {2026-01-22},
	journal = {AIP Conf. Proc.},
	author = {Dep, Linus and Vourvopoulos, George},
	month = feb,
	year = {1997},
	pages = {861--864},
}

@inproceedings{destaUseRGBImaging2017,
	title = {The use of {RGB} {Imaging} and {FTIR} {Sensors} for {Mineral} mapping in the {Reiche} {Zeche} underground test mine, {Freiberg}},
	author = {Desta, Feven and Buxton, Mike},
	month = oct,
	year = {2017},
}

@article{dehaineGeometallurgicalCharacterisationPortable2022,
	title = {Geometallurgical {Characterisation} with {Portable} {FTIR}: {Application} to {Sediment}-{Hosted} {Cu}-{Co} {Ores}},
	volume = {12},
	copyright = {http://creativecommons.org/licenses/by/3.0/},
	issn = {2075-163X},
	shorttitle = {Geometallurgical {Characterisation} with {Portable} {FTIR}},
	doi = {10.3390/min12010015},
	language = {en},
	number = {1},
	urldate = {2026-01-22},
	journal = {Minerals},
	publisher = {Multidisciplinary Digital Publishing Institute},
	author = {Dehaine, Quentin and Tijsseling, Laurens T. and Rollinson, Gavyn K. and Buxton, Mike W. N. and Glass, Hylke J.},
	month = jan,
	year = {2022},
	pages = {15},
}

@article{abdelnourPromptGammaNeutron2025,
	title = {Prompt gamma neutron activation analysis: {A} review of applications, design, analytics, challenges, and prospects},
	volume = {234},
	issn = {0969-806X},
	shorttitle = {Prompt gamma neutron activation analysis},
	doi = {10.1016/j.radphyschem.2025.112693},
	urldate = {2026-01-22},
	journal = {Radiation Physics and Chemistry},
	author = {Abdelnour, Marina R. and Liu, Juntao and Hossny, K. and Wajid, A. M. and Li, Wenxin and Liu, Zhiyi},
	month = sep,
	year = {2025},
	pages = {112693},
}

@article{niemelaRealtimeMaterialFlow2015,
	series = {4th {IFAC} {Workshop} on {Mining}, {Mineral} and {Metal} {Processing} {MMM} 2015},
	title = {Real-time {Material} {Flow} {Analysis} on {Conveyor} {Belts}},
	volume = {48},
	issn = {2405-8963},
	doi = {10.1016/j.ifacol.2015.10.071},
	number = {17},
	urldate = {2026-01-22},
	journal = {IFAC-PapersOnLine},
	author = {Niemela, A. and Hasikova, E. and Titov, V.},
	month = jan,
	year = {2015},
	pages = {24--27},
}

@article{potgieter-vermaakRamanSpectroscopyAnalysis2011,
	title = {Raman spectroscopy for the analysis of coal: a review},
	volume = {42},
	issn = {1097-4555},
	shorttitle = {Raman spectroscopy for the analysis of coal},
	doi = {10.1002/jrs.2636},
	language = {en},
	number = {2},
	urldate = {2026-01-22},
	journal = {Journal of Raman Spectroscopy},
	author = {Potgieter-Vermaak, S. and Maledi, N. and Wagner, N. and Van Heerden, J. H. P. and Van Grieken, R. and Potgieter, J. H.},
	year = {2011},
	pages = {123--129},
}

@article{uusitaloOnlineAnalysisMinerals2020,
	title = {Online analysis of minerals from sulfide ore using near-infrared {Raman} spectroscopy},
	volume = {51},
	copyright = {© 2020 The Authors. Journal of Raman Spectroscopy published by John Wiley \& Sons Ltd},
	issn = {1097-4555},
	doi = {10.1002/jrs.5859},
	language = {en},
	number = {6},
	urldate = {2026-01-22},
	journal = {Journal of Raman Spectroscopy},
	author = {Uusitalo, Sanna and Soudunsaari, Tuomas and Sumen, Juha and Haavisto, Olli and Kaartinen, Jani and Huuskonen, Jarmo and Tuikka, Aki and Rahkamaa-Tolonen, Katariina and Paaso, Janne},
	year = {2020},
	pages = {978--988},
}

@book{jobRealtimeShovelMounted2017,
	title = {Real-time shovel mounted coal or ore sensing.},
	author = {Job, Andrew and Edgar, Michael and Mcaree, Peter},
	month = jul,
	year = {2017},
}

@misc{InteriorDepartmentReleases,
	title = {Interior {Department} releases final 2025 {List} of {Critical} {Minerals} {\textbar} {U}.{S}. {Geological} {Survey}},
	language = {en},
	year = {Accessed: 2026-01-22},
}

@misc{canadaCriticalMineralsOpportunity2022,
	type = {campaigns},
	title = {Critical minerals: an opportunity for {Canada}},
	shorttitle = {Critical minerals},
	language = {eng},
	urldate = {2026-01-22},
	author = {Canada, Service},
	month = apr,
	year = {2022},
	note = {Last Modified: 2025-05-05},
}

@article{degiacomoEffectsBackgroundEnvironment2012,
	title = {Effects of the background environment on formation, evolution and emission spectra of laser-induced plasmas},
	volume = {78},
	issn = {0584-8547},
	doi = {10.1016/j.sab.2012.10.003},
	urldate = {2026-01-15},
	journal = {Spectrochimica Acta Part B: Atomic Spectroscopy},
	author = {De Giacomo, A. and Dell'Aglio, M. and Gaudiuso, R. and Amoruso, S. and De Pascale, O.},
	month = dec,
	year = {2012},
	pages = {1--19},
}

@article{diazLaserinducedBreakdownSpectroscopy2026,
	title = {Laser-induced breakdown spectroscopy for the characterization of certified reference materials containing rare earth elements},
	volume = {237},
	issn = {0584-8547},
	doi = {10.1016/j.sab.2025.107420},
	urldate = {2026-01-14},
	journal = {Spectrochimica Acta Part B: Atomic Spectroscopy},
	author = {Diaz, Daniel and Fayyaz, Amir and Baig, Muhammad Aslam and Wilson, Tyler and Hahn, David W.},
	month = mar,
	year = {2026},
	pages = {107420},
}

@article{dalmDiscriminatingOreWaste2017,
	title = {Discriminating ore and waste in a porphyry copper deposit using short-wavelength infrared ({SWIR}) hyperspectral imagery},
	volume = {105},
	issn = {0892-6875},
	doi = {10.1016/j.mineng.2016.12.013},
	urldate = {2026-01-05},
	journal = {Minerals Engineering},
	author = {Dalm, M. and Buxton, M. W. N. and van Ruitenbeek, F. J. A.},
	month = may,
	year = {2017},
	pages = {10--18},
}

@article{coghillBulkSortingTrial2024,
	title = {A bulk sorting trial of copper ore using a magnetic resonance analyser},
	volume = {210},
	issn = {0892-6875},
	doi = {10.1016/j.mineng.2024.108664},
	urldate = {2026-01-05},
	journal = {Minerals Engineering},
	author = {Coghill, Peter J. and Simpemba, Evans},
	month = may,
	year = {2024},
	pages = {108664},
}

@article{coddingtonDualcombSpectroscopy2016,
	title = {Dual-comb spectroscopy},
	volume = {3},
	issn = {2334-2536},
	doi = {10.1364/OPTICA.3.000414},
	language = {EN},
	number = {4},
	urldate = {2024-02-22},
	journal = {Optica, OPTICA},
	publisher = {Optica Publishing Group},
	author = {Coddington, Ian and Newbury, Nathan and Swann, William},
	month = apr,
	year = {2016},
	pages = {414--426},
}

@article{weeksMeasurementNeutralGadolinium2021a,
	title = {Measurement of neutral gadolinium oscillator strengths using dual-comb absorption spectroscopy in laser-produced plasmas},
	volume = {181},
	issn = {0584-8547},
	doi = {10.1016/j.sab.2021.106199},
	urldate = {2024-07-16},
	journal = {Spectrochimica Acta Part B: Atomic Spectroscopy},
	author = {Weeks, Reagan R. D. and Phillips, Mark C. and Zhang, Yu and Harilal, Sivanandan S. and Jones, R. Jason},
	month = jul,
	year = {2021},
	pages = {106199},
}

@article{weeksMultispeciesTemperatureNumber2022c,
	title = {Multi-species temperature and number density analysis of a laser-produced plasma using dual-comb spectroscopy},
	volume = {131},
	issn = {0021-8979},
	doi = {10.1063/5.0094213},
	number = {22},
	urldate = {2024-05-30},
	journal = {Journal of Applied Physics},
	author = {Weeks, Reagan R. D. and Zhang, Yu and Harilal, Sivanandan S. and Phillips, Mark C. and Jones, R. Jason},
	month = jun,
	year = {2022},
	pages = {223103},
}

@article{rhoadesDualcombAbsorptionSpectroscopy2022,
	title = {Dual-comb absorption spectroscopy of molecular {CeO} in a laser-produced plasma},
	volume = {47},
	copyright = {© 2022 Optica Publishing Group},
	issn = {1539-4794},
	doi = {10.1364/OL.455237},
	language = {EN},
	number = {10},
	urldate = {2024-06-27},
	journal = {Opt. Lett., OL},
	publisher = {Optica Publishing Group},
	author = {Rhoades, Ryan T. and Weeks, Reagan R. D. and Erickson, Seth E. and Lecaplain, Caroline and Harilal, Sivanandan S. and Phillips, Mark C. and Jones, R. Jason},
	month = may,
	year = {2022},
	pages = {2502--2505},
}

@article{mccauleyDualcombSpectroscopyDeep2024a,
	title = {Dual-comb spectroscopy in the deep ultraviolet},
	volume = {11},
	issn = {2334-2536},
	doi = {10.1364/OPTICA.516851},
	language = {EN},
	number = {4},
	urldate = {2024-04-05},
	journal = {Optica, OPTICA},
	publisher = {Optica Publishing Group},
	author = {McCauley, John J. and Phillips, Mark C. and Weeks, Reagan R. D. and Zhang, Yu and Harilal, Sivanandan S. and Jones, R. Jason},
	month = apr,
	year = {2024},
	pages = {460--463},
}

@article{westernPGOPHERProgramSimulating2017,
	series = {Satellite {Remote} {Sensing} and {Spectroscopy}: {Joint} {ACE}-{Odin} {Meeting}, {October} 2015},
	title = {{PGOPHER}: {A} program for simulating rotational, vibrational and electronic spectra},
	volume = {186},
	issn = {0022-4073},
	shorttitle = {{PGOPHER}},
	doi = {10.1016/j.jqsrt.2016.04.010},
	urldate = {2025-10-14},
	journal = {Journal of Quantitative Spectroscopy and Radiative Transfer},
	author = {Western, Colin M.},
	month = jan,
	year = {2017},
	pages = {221--242},
}

@inproceedings{jarymowyczCombingRareEarths2025,
	title = {Combing for the {Rare}-{Earths}},
	copyright = {© 2025 The Author(s)},
	doi = {10.1364/CLEO\_SI.2025.SS119\_2},
	language = {EN},
	urldate = {2025-10-02},
	booktitle = {{CLEO} 2025 (2025), paper {SS119}\_2},
	publisher = {Optica Publishing Group},
	author = {Jarymowycz, Andrew and Dannar, Hope and Hofer, Christina and McCauley, John J. and Tooley, Dylan P. and Bowman, Errol and Phillips, Mark C. and Jones, David J. and Jones, R. Jason},
	month = may,
	year = {2025},
}

@misc{kuruczAtomicLineData1995,
	title = {Atomic {Line} {Data}},
	urldate = {2025-09-19},
	journal = {Robert Kurucz CD-ROM},
	author = {Kurucz, Robert and Bell, B.},
	month = jan,
	year = {1995, Last accessed: 2026-07-06},
}

@article{harilalOpticalSpectroscopyLaserproduced2018d,
	title = {Optical spectroscopy of laser-produced plasmas for standoff isotopic analysis},
	volume = {5},
	issn = {1931-9401},
	doi = {10.1063/1.5016053},
	number = {2},
	urldate = {2025-09-19},
	journal = {Appl. Phys. Rev.},
	author = {Harilal, S. S. and Brumfield, B. E. and LaHaye, N. L. and Hartig, K. C. and Phillips, M. C.},
	month = apr,
	year = {2018},
	pages = {021301},
}

@inproceedings{bowmanDualCombSpectroscopySystem2025,
	title = {A {Dual}-{Comb} {Spectroscopy} {System} for the {Detection} of {Rare}-{Earth} {Minerals}},
	issn = {2693-8316},
	doi = {10.1109/PN66844.2025.11097153},
	urldate = {2025-09-19},
	booktitle = {2025 {Photonics} {North} ({PN})},
	author = {Bowman, Errol and Hofer, Christina and Wong, Avery and McCauley, John J. and Tooley, Dylan P. and Jarymowycz, Andrew and Dannar, Hope and Mills, Arthur K. and Phillips, Mark C. and Jones, R. Jason and Jones, David J.},
	month = may,
	year = {2025},
	pages = {1--2},
}

@article{zhangTimeresolvedDualcombMeasurement2019,
	title = {Time-resolved dual-comb measurement of number density and temperature in a laser-induced plasma},
	volume = {44},
	copyright = {© 2019 Optical Society of America},
	issn = {1539-4794},
	doi = {10.1364/OL.44.003458},
	language = {EN},
	number = {14},
	urldate = {2024-08-09},
	journal = {Opt. Lett., OL},
	publisher = {Optica Publishing Group},
	author = {Zhang, Yu and Lecaplain, Caroline and Weeks, Reagan R. D. and Yeak, Jeremy and Harilal, Sivanandan S. and Phillips, Mark C. and Jones, R. Jason},
	month = jul,
	year = {2019},
	pages = {3458--3461},
}

@misc{AtomicSpectraDatabase2009,
	title = {Atomic {Spectra} {Database}},
	language = {en},
	urldate = {2025-09-19},
	journal = {NIST},
	month = jul,
	year = {2009},
	note = {Last Modified: 2025-02-24T15:28-05:00},
}

@article{harilalSpectroscopicCharacterizationLaserinduced2005,
	title = {Spectroscopic characterization of laser-induced tin plasma},
	volume = {98},
	issn = {0021-8979},
	doi = {10.1063/1.1977200},
	number = {1},
	urldate = {2025-09-19},
	journal = {J. Appl. Phys.},
	author = {Harilal, S. S. and O’Shay, Beau and Tillack, Mark. S. and Mathew, Manoj V.},
	month = jul,
	year = {2005},
	pages = {013306},
}

@inproceedings{hoferDualCombSpectroscopyRareEarthElementDetection2025,
	title = {Dual-{Comb}-{Spectroscopy} for {Rare}-{Earth}-{Element} {Detection}},
	issn = {2833-1052},
	doi = {10.1109/CLEO/Europe-EQEC65582.2025.11111108},
	urldate = {2025-09-19},
	booktitle = {2025 {Conference} on {Lasers} and {Electro}-{Optics} {Europe} \& {European} {Quantum} {Electronics} {Conference} ({CLEO}/{Europe}-{EQEC})},
	author = {Hofer, Christina and Jarymowycz, Andrew and Dannar, Hope and McCauley, John J. and Bowman, Errol and Tooley, Dylan P. and Wong, Avery and Mills, Arthur K. and Phillips, Mark C. and Jones, R. Jason and Jones, David J.},
	month = jun,
	year = {2025},
	pages = {1--1},
}

@article{martinQuantificationRareEarth2015,
	title = {Quantification of rare earth elements using laser-induced breakdown spectroscopy},
	volume = {114},
	issn = {0584-8547},
	doi = {10.1016/j.sab.2015.10.005},
	urldate = {2025-10-27},
	journal = {Spectrochimica Acta Part B: Atomic Spectroscopy},
	author = {Martin, Madhavi and Martin, Rodger C. and Allman, Steve and Brice, Deanne and Wymore, Ann and Andre, Nicolas},
	month = dec,
	year = {2015},
	pages = {65--73},
}

@article{gondalRoleVariousBinding2007a,
	title = {The role of various binding materials for trace elemental analysis of powder samples using laser-induced breakdown spectroscopy},
	volume = {72},
	issn = {0039-9140},
	doi = {10.1016/j.talanta.2006.11.039},
	number = {2},
	urldate = {2025-10-22},
	journal = {Talanta},
	author = {Gondal, M. A. and Hussain, T. and Yamani, Z. H. and Baig, M. A.},
	month = apr,
	year = {2007},
	pages = {642--649},
}

@article{phillipsDetectionLimitsLaser2025,
	title = {Detection limits for laser absorption spectroscopy of {Li} in laser ablation plumes},
	volume = {50},
	issn = {0146-9592, 1539-4794},
	doi = {10.1364/OL.564323},
	language = {en},
	number = {10},
	urldate = {2025-10-20},
	journal = {Opt. Lett.},
	author = {Phillips, Mark C. and Kautz, Elizabeth J. and Harilal, Sivanandan S.},
	month = may,
	year = {2025},
	pages = {3349},
}

@article{harilalOpticalDiagnosticsLaserproduced2022,
	title = {Optical diagnostics of laser-produced plasmas},
	volume = {94},
	doi = {10.1103/RevModPhys.94.035002},
	number = {3},
	journal = {Rev. Mod. Phys.},
	author = {Harilal, S. S.},
	year = {2022},
}

@article{bhattDeterminationRareEarth2018,
	title = {Determination of {Rare} {Earth} {Elements} in {Geological} {Samples} {Using} {Laser}-{Induced} {Breakdown} {Spectroscopy} ({LIBS})},
	volume = {72},
	issn = {0003-7028},
	doi = {10.1177/0003702817734854},
	language = {EN},
	number = {1},
	urldate = {2025-10-24},
	journal = {Appl Spectrosc},
	publisher = {SAGE Publications Ltd STM},
	author = {Bhatt, Chet R. and Jain, Jinesh C. and Goueguel, Christian L. and McIntyre, Dustin L. and Singh, Jagdish P.},
	month = jan,
	year = {2018},
	pages = {114--121},
}

@article{bernathStypeStarsLine2023,
	title = {S-type {Stars}: {Line} {List} for the {A2Π}–{X2Σ}+ {Band} {System} of {LaO}},
	volume = {953},
	issn = {0004-637X},
	shorttitle = {S-type {Stars}},
	doi = {10.3847/1538-4357/ace68a},
	language = {en},
	number = {2},
	urldate = {2025-10-14},
	journal = {ApJ},
	publisher = {The American Astronomical Society},
	author = {Bernath, P. F. and Dodangodage, R. and Liévin, J.},
	month = aug,
	year = {2023},
	pages = {181},
}

@article{muravievDualfrequencycombUVSpectroscopy2024,
	title = {Dual-frequency-comb {UV} spectroscopy with one million resolved comb lines},
	volume = {11},
	copyright = {\&\#169; 2024 Optica Publishing Group},
	issn = {2334-2536},
	doi = {10.1364/OPTICA.536971},
	language = {EN},
	number = {11},
	urldate = {2025-02-04},
	journal = {Optica, OPTICA},
	publisher = {Optica Publishing Group},
	author = {Muraviev, Andrey and Konnov, Dmitrii and Vasilyev, Sergey and Vodopyanov, Konstantin L.},
	month = nov,
	year = {2024},
	pages = {1486--1489},
}

@article{newburySensitivityCoherentDualcomb2010,
	title = {Sensitivity of coherent dual-comb spectroscopy},
	volume = {18},
	copyright = {\&\#169; 2010 OSA},
	issn = {1094-4087},
	doi = {10.1364/OE.18.007929},
	language = {EN},
	number = {8},
	urldate = {2023-11-16},
	journal = {Opt. Express, OE},
	publisher = {Optica Publishing Group},
	author = {Newbury, Nathan R. and Coddington, Ian and Swann, William},
	month = apr,
	year = {2010},
	pages = {7929--7945},
}

@article{bergevinDualcombSpectroscopyLaserinduced2018b,
	title = {Dual-comb spectroscopy of laser-induced plasmas},
	volume = {9},
	issn = {2041-1723},
	doi = {10.1038/s41467-018-03703-0},
	language = {en},
	number = {1},
	urldate = {2023-02-14},
	journal = {Nat Commun},
	author = {Bergevin, Jenna and Wu, Tsung-Han and Yeak, Jeremy and Brumfield, Brian E. and Harilal, Sivanandan S. and Phillips, Mark C. and Jones, R. Jason},
	month = mar,
	year = {2018},
	pages = {1273},
}

@article{zawiszaDeterminationRareEarth2011,
	title = {Determination of rare earth elements by spectroscopic techniques: a review},
	volume = {26},
	issn = {0267-9477, 1364-5544},
	shorttitle = {Determination of rare earth elements by spectroscopic techniques},
	doi = {10.1039/c1ja10140d},
	language = {en},
	number = {12},
	urldate = {2026-06-18},
	journal = {J. Anal. At. Spectrom.},
	author = {Zawisza, Beata and Pytlakowska, Katarzyna and Feist, Barbara and Polowniak, Marzena and Kita, Andrzej and Sitko, Rafal},
	year = {2011},
	pages = {2373},
}

@book{welzHighResolutionContinuumSource2005,
	edition = {1},
	title = {High‐{Resolution} {Continuum} {Source} {AAS}: {The} {Better} {Way} to {Do} {Atomic} {Absorption} {Spectrometry}},
	copyright = {http://doi.wiley.com/10.1002/tdm\_license\_1.1},
	isbn = {978-3-527-30736-4 978-3-527-60651-1},
	shorttitle = {High‐{Resolution} {Continuum} {Source} {AAS}},
	doi = {10.1002/3527606513},
	language = {en},
	urldate = {2026-06-18},
	publisher = {Wiley},
	author = {Welz, Bernhard and Becker‐Ross, Helmut and Florek, Stefan and Heitmann, Uwe},
	month = jan,
	year = {2005},
}

@article{gaftImagingRareearthElements2019a,
	title = {Imaging rare-earth elements in minerals by laser-induced plasma spectroscopy: {Molecular} emission and plasma-induced luminescence},
	volume = {151},
	issn = {0584-8547},
	shorttitle = {Imaging rare-earth elements in minerals by laser-induced plasma spectroscopy},
	doi = {10.1016/j.sab.2018.11.003},
	urldate = {2026-06-25},
	journal = {Spectrochimica Acta Part B: Atomic Spectroscopy},
	author = {Gaft, M. and Raichlin, Y. and Pelascini, F. and Panzer, G. and Motto Ros, V.},
	month = jan,
	year = {2019},
	pages = {12--19},
}

@article{duttOnchipDualcombSource2018,
	title = {On-chip dual-comb source for spectroscopy},
	volume = {4},
	doi = {10.1126/sciadv.1701858},
	number = {3},
	urldate = {2026-06-25},
	journal = {Science Advances},
	publisher = {American Association for the Advancement of Science},
	author = {Dutt, Avik and Joshi, Chaitanya and Ji, Xingchen and Cardenas, Jaime and Okawachi, Yoshitomo and Luke, Kevin and Gaeta, Alexander L. and Lipson, Michal},
	month = mar,
	year = {2018},
	pages = {e1701858},
}

@article{zhongBroadbandPhotoncountingDualcomb2025,
	title = {Broadband photon-counting dual-comb spectroscopy with attowatt sensitivity over turbulent optical paths},
	volume = {14},
	copyright = {2025 The Author(s)},
	issn = {2047-7538},
	doi = {10.1038/s41377-025-01934-7},
	language = {en},
	number = {1},
	urldate = {2026-06-25},
	journal = {Light Sci Appl},
	publisher = {Nature Publishing Group},
	author = {Zhong, Wei and Liu, Yingyu and Yin, Qin and Zhao, Ruocan and Wang, Chong and Ren, Wei and Dou, Xiankang and Xue, Xianghui},
	month = aug,
	year = {2025},
	pages = {293},
}

@article{hermanPreciseMultispeciesAgricultural2021,
	title = {Precise multispecies agricultural gas flux determined using broadband open-path dual-comb spectroscopy},
	volume = {7},
	doi = {10.1126/sciadv.abe9765},
	number = {14},
	urldate = {2026-06-25},
	journal = {Science Advances},
	publisher = {American Association for the Advancement of Science},
	author = {Herman, Daniel I. and Weerasekara, Chinthaka and Hutcherson, Lindsay C. and Giorgetta, Fabrizio R. and Cossel, Kevin C. and Waxman, Eleanor M. and Colacion, Gabriel M. and Newbury, Nathan R. and Welch, Stephen M. and DePaola, Brett D. and Coddington, Ian and Santos, Eduardo A. and Washburn, Brian R.},
	month = mar,
	year = {2021},
	pages = {eabe9765},
}

@phdthesis{weeksDualCombSpectroscopyLaserProduced2023,
	address = {United States -- Arizona},
	type = {Ph.{D}.},
	title = {Dual-{Comb} {Spectroscopy} of {Laser}-{Produced} {Plasmas}},
	copyright = {Database copyright ProQuest LLC; ProQuest does not claim copyright in the individual underlying works.},
	isbn = {979-8-3795-3491-2},
	language = {English},
	urldate = {2023-09-05},
	school = {The University of Arizona},
	author = {Weeks, Reagan R. D.},
	year = {2023},
}

@article{phillips2026CalibrationFree,
	title = {Calibration-{Free} {Analysis} of {Li} {Isotope} {Ratios} {Using} {Laser} {Ablation} and {Laser} {Absorption} {Spectroscopy}},
	volume = {98},
	copyright = {https://doi.org/10.15223/policy-029},
	issn = {0003-2700, 1520-6882},
	doi = {10.1021/acs.analchem.6c01718},
	language = {en},
	number = {25},
	urldate = {2026-07-05},
	journal = {Anal. Chem.},
	author = {Phillips, Mark C. and Makovsky, Kyle A. and Stevens, Richard E. and Kautz, Elizabeth J. and Harilal, Sivanandan S.},
	month = jun,
	year = {2026},
	pages = {18873--18881},
}
\end{document}